\documentclass[journal=jacsat,manuscript=article]{achemso}

\usepackage{chemformula} 
\usepackage[T1]{fontenc} 

\usepackage{verbatim}
\usepackage{hyperref}
\usepackage{mathtools}
\usepackage{wrapfig}
\usepackage{algorithm}
\usepackage{algpseudocode}
\usepackage{graphicx}

\usepackage{amsfonts}

\newcommand{\bs}{\boldsymbol}

\newcommand{\bveps}{{\bs \varepsilon}}

\RequirePackage{xspace}
\makeatletter
\DeclareRobustCommand\onedot{\futurelet\@let@token\@onedot}
\def\@onedot{\ifx\@let@token.\else.\null\fi\xspace}

\def\eg{\emph{e.g}\onedot}
\def\ie{\emph{i.e}\onedot}
\makeatother

\author{Dongjin Seo}
\altaffiliation{These authors contributed equally to this work.}
\affiliation[Yale University]
{Department of Applied Physics, Yale University, New Haven, Connecticut 06511, USA}

\author{Soobin Um}
\altaffiliation{These authors contributed equally to this work.}
\affiliation[KAIST]
{Graduate School of AI, KAIST, Daejeon 34141, Republic of Korea}

\author{Sangbin Lee}
\affiliation[Hanyang University]
{Department of Electrical Engineering, Hanyang University, Seoul 04763, Republic of Korea}

\author{Jong Chul Ye}
\affiliation[KAIST]
{Graduate School of AI, KAIST, Daejeon 34141, Republic of Korea}
\email{jong.ye@kaist.ac.kr}

\author{Haejun Chung}
\affiliation[Hanyang University]
{Department of Electrical Engineering, Hanyang University, Seoul 04763, Republic of Korea}
\email{haejun@hanyang.ac.kr}

\title{Physics-guided and fabrication-aware\\inverse design of photonic devices\\using diffusion models}

\abbreviations{IR,NMR,UV}
\keywords{American Chemical Society, \LaTeX}

\begin{document}

\begin{tocentry}





\centering
\includegraphics[height=4cm,keepaspectratio]{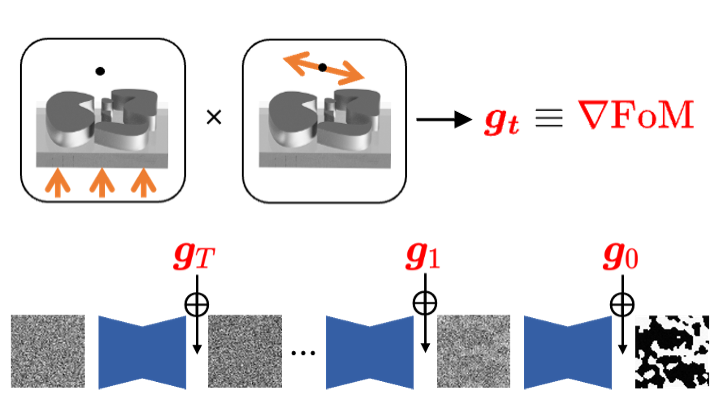}

\end{tocentry}

\begin{abstract}

Designing free‑form photonic devices is fundamentally challenging due to the vast number of possible geometries and the complex requirements of fabrication constraints. Traditional inverse‑design approaches—whether driven by human intuition, global optimization, or adjoint‑based gradient methods—often involve intricate binarization and filtering steps, while recent deep‑learning strategies demand prohibitively large numbers of simulations ($10^5$-$10^6$). To overcome these limitations, we present \emph{AdjointDiffusion}, a physics-guided framework that integrates adjoint sensitivity gradients into the sampling process of diffusion models.
\emph{AdjointDiffusion} begins by training a diffusion network on a synthetic, fabrication‑aware dataset of binary masks. During inference, we compute the adjoint gradient of a candidate structure and inject this physics‑based guidance at each denoising step, steering the generative process toward high figure‑of‑merit (FoM) solutions without additional post‑processing. We demonstrate our method on two canonical photonic design problems—a bent waveguide and a CMOS image sensor color router—and show that our method consistently outperforms state‑of‑the‑art nonlinear optimizers (\eg, MMA, SLSQP) in both efficiency and manufacturability, while using orders of magnitude fewer simulations ($\sim 2 \times 10^2$) than pure deep‑learning approaches ($\sim10^5$–$10^6$).
By eliminating complex binarization schedules and minimizing simulation overhead, \emph{AdjointDiffusion} offers a streamlined, simulation‑efficient, and fabrication‑aware pipeline for next‑generation photonic device design. Our open‑source implementation is available at \href{https://github.com/dongjin-seo2020/AdjointDiffusion}{https://github.com/dongjin-seo2020/AdjointDiffusion}.

\end{abstract}

* Preliminary results of this study were presented at [2024 Conference on Lasers and Electro-Optics Pacific Rim (CLEO-PR) (proceeding)]~\cite{10676695} and [NeurIPS ML4PS Workshop 2024 (non-archival)]~\cite{seophysics}.

\section{Introduction}

Photonic device design offers the control and manipulation of light in a wide array of applications --- from metasurfaces~\cite{} and photonic integrated circuits~\cite{} to quantum photonics~\cite{} and beyond~\cite{}. Early design efforts were guided primarily by human intuition~\cite{bermel2007improving, tsakalakos2007silicon, yu2014flat, chen2018broadband}, yielding simple geometries whose limited complexity often constrained device functionality. With increasing demand for multifunctional, high‐performance components, free form topologies have emerged as a powerful alternative, enabling exploration of vastly larger design spaces and unlocking superior performance~\cite{piggott2015inverse, chung2020high, kim2024freeform, roberts20233d}.

To navigate these high dimensional spaces efficiently, inverse design techniques—particularly those based on adjoint sensitivity analysis—have gained prominence~\cite{molesky2018inverse, christiansen2021inverse, Miller:EECS-2012-115, CAO2002171, lalau2013adjoint, allaire01242950}. By running just two electromagnetic simulations per iteration (one “forward” and one “adjoint”), these methods obtain exact gradients of a Figure of Merit (FoM) with respect to each design parameter. In practice, however, they demand intricate filtering and binarization routines to enforce fabrication constraints and are susceptible to convergence in suboptimal local minima.

Complementary to gradient-based solvers, deep learning frameworks --- including generative models~\cite{jiang2019free, jiang2019global} and reinforcement learning (RL) agents~\cite{sajedian2019optimisation, seoRL, park2024sample} --- treat inverse design from a data-driven perspective. Generative networks translate device layout design into an image generation task, while RL policies iteratively improve a design through simulated interactions. Despite their flexibility, these approaches typically require the number of simulations on the order of $10^5$–$10^6$, limiting their practicality for complex photonic problems.

In this work, we introduce \emph{AdjointDiffusion}, a physics-guided inverse design framework that integrates adjoint sensitivity analysis with denoising diffusion probabilistic models~\cite{song2019generative, ho2020denoising}. We begin by constructing a synthetic, fabrication-aware dataset of binary structures --- each smoothed via Gaussian filtering and re-binarized --- and train a conditional diffusion model to capture this fabrication-compliant prior. During inference, we generate the final structure by running the reverse diffusion process. We then compute how the FoM changes with respect to this final design, and use that gradient to guide each step of the reverse process. This guided sampling process naturally drives the generation of high-performance, fabrication-ready layouts without the need for elaborate post-processing or extensive hyperparameter tuning.
Although diffusion models have recently been applied to photonic inverse design~\cite{Zhang_2023, yang2024guided, vlastelica2023diffusion}, none have embedded adjoint gradients into the sampling loop. Also, diffusion models for inverse design typically require a large amount of simulation data for training (on the order of $\sim10^5$ samples~\cite{Zhang_2023}). We circumvent this limitation by training on binary data and relying on simulations only during the inference phase.  We demonstrate our method on two canonical problems—a bent waveguide and a CMOS image sensor color router—and show that it achieves superior or competitive FoM using only on the order of $10^2$  simulations. \textit{AdjointDiffusion} thus offers a simulation-efficient and fabrication-aware pipeline for next-generation photonic device design.

\begin{figure}[t]
\centering
  \includegraphics[width=\textwidth]{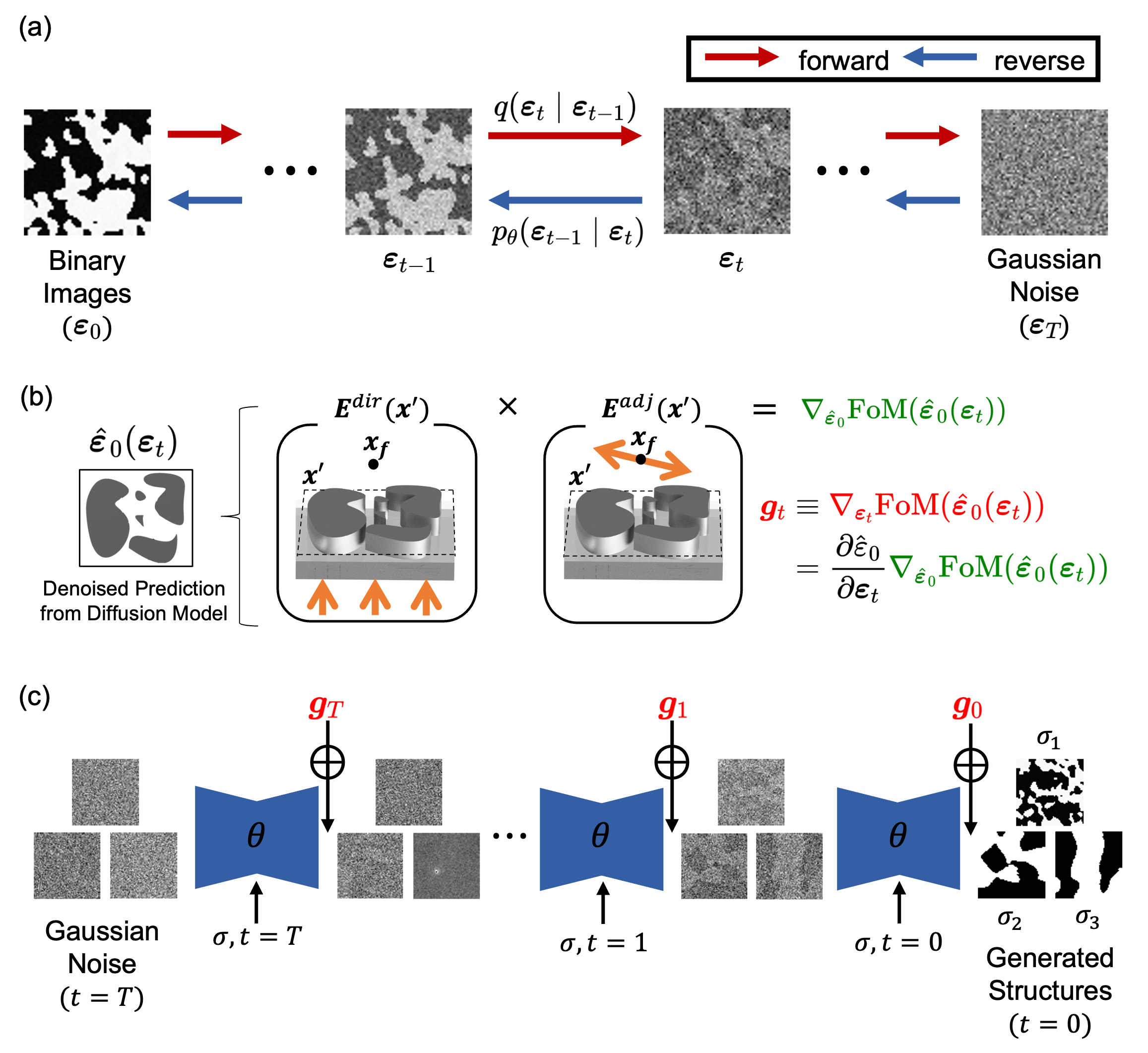}

  \caption{
  Illustration of the process of integrating adjoint optimization with diffusion models for generating structures. (a) The forward and reverse diffusion process. The forward process (red arrows) starts from binary images ($\hat{\mathbf{\varepsilon}}_0$) and adds Gaussian noise until the images are completely noisy. The reverse process (blue arrows) denoises the noisy images step-by-step, using a learned model $\theta$ to reconstruct the binary images.
   (b) Schematics of adjoint sensitivity analysis with a two-dimensional lens. The adjoint gradient $\boldsymbol{g}_t$ is calculated for the denoised prediction $\hat{\boldsymbol{\varepsilon}}_0(\boldsymbol{\varepsilon}_t)$ by the component-wise product of direct and adjoint fields. (c) The calculated gradient from (b) is added component-wise to the generated structures in the reverse process, where conditional parameters $\sigma$ and $t$ are applied.}
    \label{schematics}
\end{figure}

\section{Method}
\label{sec:method}

Before explaining the details of our work, we provide a brief overview of diffusion models to facilitate understanding of our key contributions. See the Supporting Information (S1.2) for a rigorous coverage on diffusion models.

\noindent \textbf{Notation.} Variables in bold denote matrices or vectors, while variables in regular font represent scalars. Note that our notation, where $\bveps$ represents the design parameter (i.e., relative permittivity), $\mathbf{x}$ represents the design space, and $\mathbf{z}$ represents random noise, differs from the convention commonly used in articles on diffusion models.

\subsection{Overview of diffusion-based generative models}

Diffusion models (Figure~\ref{schematics}(a)) are generative frameworks that create data through an iterative refinement process~\cite{ho2020denoising, song2019generative}. These models begin with a random noise sample and iteratively transform it into a structured output, such as an image. Formally, let $\bveps_0 \in \mathbb{R}^d$ represent a sample from the $d$-dimensional data space, drawn from the data distribution $p_{\text{data}}$, which we denote as  $\bveps_0 \sim p_{\text{data}}$. The generation process of diffusion models typically begins with random Gaussian noise $\bveps_T \sim {\cal N}( {\bf 0}, {\bf I} )$, where $\mathbf{0}$ is the zero vector and $\mathbf{I}$ is the identity matrix,  and progressively denoises it through intermediate steps $\bveps_{T-1}, \bveps_{T-2}, \dots$, until it converges to an output image $\bveps_0$, corresponding to timestep 0. Here, each state $\bveps_t \in \mathbb{R}^d$ represents a $d$-dimensional sample at time step $t$ where $t \in \{1, \dots, T\}$. The denoising process is performed by a neural network—typically a U-Net architecture~\citep{ronneberger2015u}—which is trained to infer a less noisy sample $\bveps_{t-1}$ from a noisier input $\bveps_t$. This denoising process is governed by a Gaussian transition model:
 $p_{\bs \theta} ( \bveps_{t-1} | \bveps_t ) \coloneqq {\cal N} (\bveps_{t-1}; {\bs \mu}_{\bs \theta}( \bveps_t, t ), {\bs \Sigma}_{\bs \theta}( \bveps_t, t ))$ where ${\bs \mu}_{\bs \theta}( \bveps_t, t )$ and ${\bs \Sigma}_{\bs \theta}( \bveps_t, t )$ correspond to the mean and variance governing the transition from $\bveps_t$ to $\bveps_{t-1}$~\citep{sohl2015deep, ho2020denoising}. The mean ${\bs \mu}_{\bs \theta}( \bveps_t, t )$ provides an estimate of the denoised data at step $t-1$ by subtracting the predicted noise component from $\bveps_t$, ensuring that the sample moves closer to the original data distribution. The variance ${\bs \Sigma}_{\bs \theta}( \bveps_t, t )$ accounts for the remaining uncertainty in the estimation and determines the stochasticity of the transition. This iterative refinement can be understood as a stochastic relaxation process, where the system gradually removes noise while maintaining probabilistic consistency with the original data. See Supporting Information (S1.2) for detailed mathematical expressions. 
 
Concretely, at each step $t$, the transition from $\bveps_t$ to $\bveps_{t-1}$ is performed according to:
\begin{align}
\label{eq:inference}
    \bveps_{t-1} = {\bs \mu}_{\bs \theta} ( \bveps_t, t ) + {\bs \Sigma}_{\bs \theta}^{1/2} ( \bveps_t, t ){\bf z}, \;\; {\bf z} \sim {\cal N}( {\bf 0}, {\bf I} ).
\end{align}
Given a well-trained diffusion model (\ie, trained until convergence), it is theoretically guaranteed that iterating this process~\eqref{eq:inference} from $t = T$ down to $t = 1$ reconstructs samples that are statistically close to the original data distribution, \ie, $p_{\bs \theta} \approx p_{\text{data}}$.

\noindent \textbf{Conditional generation via guidance.} A key advantage of diffusion-based samplers is that they admit interventions in the iterative denoising process to implement conditional generation~\citep{dhariwal2021diffusion}. Specifically, given a differentiable target utility function $f(\boldsymbol{\varepsilon}_t,\,t,\,\mathbf{c})$ that may depend on an arbitrary condition ${\bf c}$, one can incorporate its gradient into the update equation~\eqref{eq:inference} to yield a \emph{guided} sampler:
\begin{equation}
\label{eq:guided_inference_conditional}
\boldsymbol{\varepsilon}_{t-1} = \boldsymbol{\mu}_{\boldsymbol{\theta}}(\boldsymbol{\varepsilon}_t, t) + \boldsymbol{\Sigma}_{\boldsymbol{\theta}}^{1/2}(\boldsymbol{\varepsilon}_t, t)\,\mathbf{z} + \eta_t\,\nabla_{\boldsymbol{\varepsilon}_t} f(\boldsymbol{\varepsilon}_t, t, \mathbf{c}), \quad \mathbf{z} \sim \mathcal{N}(\mathbf{0}, \mathbf{I}).
\end{equation}

where $\eta_t$ corresponds to the strength of guidance term $\nabla { f}$, which may be scheduled over inference step $t$. A well-known example is classifier guidance~\citep{dhariwal2021diffusion}, where the log-likelihood of a noise-aware class predictor is incorporated: ${f}(\bveps_t, t, y) \coloneqq \log p_{\bs \phi}( y | \bveps_t )$. Here, ${\bs \phi}$ represents the parameters of the classifier, and $y$ denotes the selected condition for the generated output. 

Although effective, many guided samplers, including classifier guidance, require potentially expensive additional resources like data collection and model training. 
This requirement limits the applicability of these techniques in challenging scenarios, such as our focused inverse design problem where obtaining a large volume of high-quality structural data is prohibitive.
In the next section, we develop a \emph{data-free} framework for inverse design by intelligently incorporating adjoint sensitivity analysis into the concept of guided sampling, which offers significant performance improvements over traditional inverse design solvers with minimal implementation complexities.

\subsection{AdjointDiffusion: physics-guided diffusion models for inverse design}
\label{PGDM}

We propose our physics-guided diffusion model algorithm named \textit{AdjointDiffusion}, which combines the adjoint sensitivity analysis with the generation pipeline of diffusion models.
Our approach consists of four distinct stages: (i) dataset construction, (ii) diffusion model training, (iii)  structure generation, and (iv) post-processing.



\noindent \textbf{(i) Dataset construction.} We first prepare training data for constructing a diffusion model (\ie, a denoising network $p_{\bs \theta}$) specifically designed for our inverse design framework.
To generate structured training data, we first define a design parameter ranging from zero to one, where zero means a material with lower permittivity and one means a material with higher permittivity. Then, we start by creating 30,000 random binary (design parameter with zero or one) data with a resolution of $64 \times 64$ as shown in Figure~\ref{schematics}(a).

To incorporate fabrication constraints into the dataset, we apply Gaussian filters to the randomly generated structures, which smooth the sharp edges of the material distribution in space, resulting in a design parameter distribution with diverse density values ranging from 0 to 1~\cite{hidaka2021topology}. 
The dataset is evenly divided into three subsets of 10,000 images, each processed with a Gaussian filter of standard deviation 2, 5, or 8.
The filtered structures are then binarized using a 0.5 threshold, enforcing a two-material (0 and 1) composition.
With this approach, we enable the conditional generation of structures within a single network, allowing it to accommodate multiple fabrication constraints while being trained only once.
Also, since our algorithm can inherently work with any dataset by simply training the diffusion model on it, one can apply their datasets following different designs and applications.

\noindent \textbf{(ii) Diffusion model training.} With the generated training dataset, we subsequently train a diffusion model $p_{\bs \theta}$ following the methodology proposed by~\citet{dhariwal2021diffusion}. Specifically, we use $1,000$ timesteps (i.e., $T=1000$) with the cosine noise scheduling~\citep{nichol2021improved}. For the training objective, we adopt the hybrid loss $L_{\text{hybrid}}$ introduced by~\citet{nichol2021improved}. For each training sample, the associated fabrication constraint (\ie, the Gaussian filter's standard deviation) is incorporated as a conditioning parameter $\sigma$, where the associated Gaussian transition is then expressible as: $p_{\bs \theta} ( \bveps_{t-1} | \bveps_t, \sigma ) \coloneqq {\cal N} (\bveps_{t-1}; {\bs \mu}_{\bs \theta}( \bveps_t, t, \sigma ), {\bs \Sigma}_{\bs \theta}( \bveps_t, t, \sigma ))$. This training enables the diffusion model to generate binary data that meet the specified fabrication conditions $\sigma$. See the Supporting Information (S1.2 and S2) for details on the training process.

\noindent \textbf{(iii) Structure generation.} The third stage, which forms the core of our framework, involves generating high-performance photonic structures with enhanced FoM. One straightforward approach to achieve this is to select a desired feature size $\sigma$ and generate binary structures by applying the following equation (iteratively from $t=T$ down to $t=1$):
\begin{align}
\label{eq:inference_sigma}
    \bveps_{t-1} = {\bs \mu}_{\bs \theta} ( \bveps_t, t, \sigma ) + {\bs \Sigma}_{\bs \theta}^{1/2} ( \bveps_t, t, \sigma ){\bf z}. 
\end{align}
However, as demonstrated in our results (see Supporting Information S3), this approach does not yield high-FoM structures with high FoM values. An alternative is to use a diffusion model trained on high-FoM structural data for sampling, but as previously mentioned, this approach is resource-intensive due to the difficulty of obtaining such high-quality data.

We make alteration by leveraging the principle of guided sampling in diffusion models. Specifically, we intervene in the sampling process~\eqref{eq:inference_sigma} and incorporate the FoM-maximizing direction, calculated using the adjoint sensitivity analysis, to guide the generation process toward high-FoM regions of the data space.

To achieve this, we first compute the posterior mean $\hat{\bs{\varepsilon}}_0(\bs{\varepsilon}_t, \sigma) \coloneqq \mathbb{E}[\bveps_0 | \bveps_t, \sigma]$ by employing the pretrained diffusion model and Tweedie's formula~\citep{robbins1992empirical} (see Supporting Information S1.2 for details on the formula). Next, we calculate the adjoint gradient of the predicted structure, $\nabla_{\hat{\bs{\varepsilon}}_0} \text{FoM}( \hat{\bs{\varepsilon}}_0(\bs{\varepsilon}_t, \sigma) ) $, using direct and adjoint simulations~\cite{Miller:EECS-2012-115, CAO2002171} (see Figure~\ref{schematics}(b)). By computing the Jacobian $\partial \hat{\bs{\varepsilon}}_0 / \partial \bs{\varepsilon}_t$ and taking the inner product with the adjoint gradient $\nabla_{\hat{\bs{\varepsilon}}_0} \text{FoM}( \hat{\bs{\varepsilon}}_0(\bs{\varepsilon}_t, \sigma) ) $, we obtain the proposed guidance term:
\begin{align*}
    \nabla_{\bs{\varepsilon}_t} \text{FoM}( \hat{\bs{\varepsilon}}_0(\bs{\varepsilon}_t, \sigma) ) = \frac{\partial \hat{\bs{\varepsilon}}_0}{ \partial \bs{\varepsilon}_t } \nabla_{\hat{\bs{\varepsilon}}_0} \text{FoM}( \hat{\bs{\varepsilon}}_0(\bs{\varepsilon}_t, \sigma) ).
\end{align*}
See Figure~\ref{schematics}(b) for illustration.
The proposed sampler is the one that integrates this guidance into the standard sampling process~\eqref{eq:inference_sigma}, as formally expressed by:
\begin{align}
\label{eq:guided_sampling}
    \bveps_{t-1} = {\bs \mu}_{\bs \theta} ( \bveps_t, t, \sigma ) + {\bs \Sigma}_{\bs \theta}^{1/2} ( \bveps_t, t, \sigma ){\bf z} + \eta_t \nabla_{\bs{\varepsilon}_t} \text{FoM}( \hat{\bs{\varepsilon}}_0(\bs{\varepsilon}_t, \sigma) ) , 
\end{align}
where $\eta_t$ represents the strength of guidance term possibly scheduled over time $t$. Note that the update aims to optimize intermediate structures $\bveps_t$ by incorporating $\nabla_{\bs{\varepsilon}_t} \text{FoM}( \hat{\bs{\varepsilon}}_0(\bs{\varepsilon}_t, \sigma) )$ instead of $\nabla_{\bs{\varepsilon}_t} \text{FoM}(\bs{\varepsilon}_t)$. This choice of guidance often provides an advantageous aspect called \emph{manifold-constrained} property~\citep{chung2022improving, chungdiffusion, um2024self}.

Specifically, the Jacobian $\partial \hat{\bs{\varepsilon}}_0 / \partial \bs{\varepsilon}_t$, which appears in the calculation of the guidance term, can be interpreted as a \emph{projection} onto the tangent space of the data manifold ${\cal M}$ at $\hat{\bveps}_0$, denoted ${\cal T}_{\hat{\bveps}_0}{\cal M}$. This makes the guidance term point in the tangent direction relative to ${\cal M}$, thereby \emph{constraining} the updated structures to reside on the data \emph{manifold}. The manifold-constrained property offers significant empirical benefits. For instance, it naturally encourages binarization of generated structures during the generation process. This is a crucial advantage, since it eliminates the need for the binarization process, which requires tuning numerous hyperparameters and is challenging to optimize. See Figure~\ref{pixeldist} for experimental results demonstrating this advantage.



\noindent \textbf{(iv) Post-processing.} We find out that our sampler benefits from incorporating a simple post-processing pipeline on the final generated structure ${\bs \varepsilon}_0$. The post-processing consists of two stages: (1) final binarization and (2) island deletion. Firstly, since our algorithm does not use conic filters~\cite{kim2025freeform}, which act as a blurring kernel to smooth the design variables, there is a possibility that a pixel with an adjoint gradient value has an opposite sign from nearby pixels. This issue can degrade the fabricability due to the low minimum feature size (MFS). The problem is also observed in nonlinear algorithms typically used in adjoint-based optimizations, such as those provided in \textit{NLopt}~\citep{NLopt}. To address this, we apply post-processing after binarization across all algorithms. Additionally, we analyze the FoM change resulting from our post-processing in Supporting Information (S4).

We highlight several advantages of the proposed inverse design framework. First, it operates in a data-free manner, allowing the synthesis of high-quality structures without requiring real-world data. Second, our framework exhibits minimal reliance on hyperparameters. By employing a diffusion model trained on binary data, the sampling process directly generates binary structures, eliminating the need for complex binarization schedules and extensive hyperparameter tuning. Finally, our algorithm demonstrates high computational efficiency, requiring significantly fewer adjoint simulations ($2 \times 10^2$) compared to existing deep learning methods.

\section{Results and discussion}

We evaluate the performance of \textit{AdjointDiffusion} in comparison to conventional nonlinear optimization algorithms widely used in adjoint optimizations, such as MMA~\citep{MMA} and SLSQP~\citep{SLSQP0, SLSQP}; details of these algorithms are provided in Supporting Information (S1.1). Due to the stochastic nature of our algorithm, we perform three independent optimization runs using different random seeds to demonstrate robustness and reproducibility; each run generates different structures, as shown in Supporting Information (S3.4), offering diverse design options. 

\begin{figure}[t]
  \centering
  \includegraphics[width=0.5\textwidth]{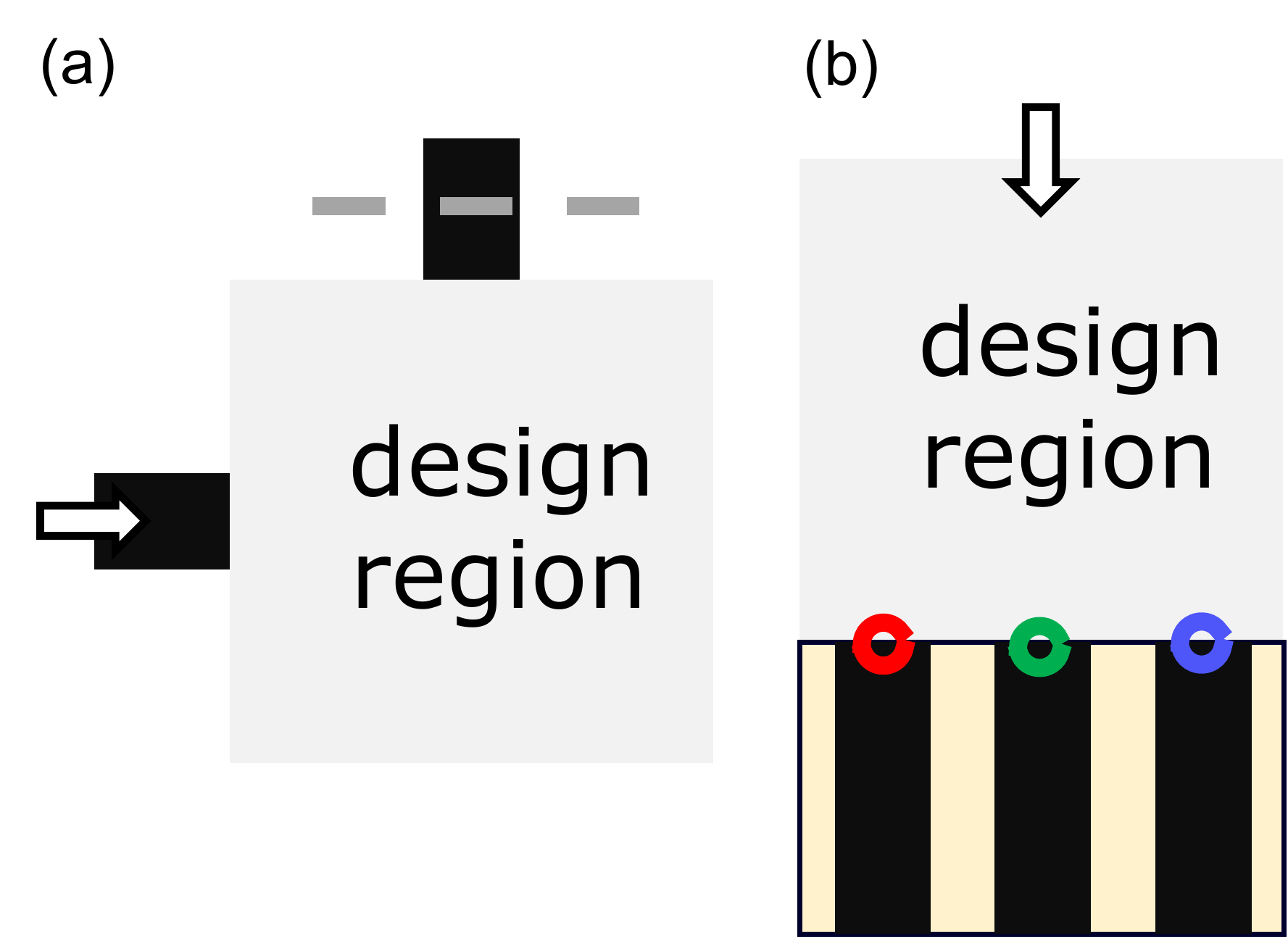}
  \caption{Schematic of the problem setup for photonic inverse design. (a) Waveguide configuration with a design region of 64 $\times$ 64 pixels degree of freedom. The white arrow represents the mono-wavelength source, while the dotted lines indicate monitors. (b) CMOS image sensor (CIS) configuration with a design region of 64 $\times$ 64 pixels degree of freedom. Bloch boundary conditions on the left and right, and a PML at the top. Colored circles indicate monitor locations for each color (R,G,B).}
  \label{problem setup}
  \end{figure}

\subsection{Preliminaries}

We use two types of simulation to validate our algorithm, as illustrated in Figure~\ref{problem setup}. The first setup (a) represents a waveguide with perfectly matched layers (PMLs) at the boundaries. The second setup (b) corresponds to a color routing-based CMOS image sensor (CIS) configuration with Bloch boundary conditions on the left and right and a PML at the top, adapted from~\citet{lee2024inverse}.

\noindent \textbf{Number of simulations.}
From the viewpoint of computational efficiency, the number of simulations required for the algorithm is an important factor. Adjoint sensitivity analysis performs two (direct- and adjoint-) simulations to calculate the adjoint gradient of the generated structure, resulting in an overall simulation requirement that is twice the number of optimization steps. We note that the reverse process of \textit{AdjointDiffusion} can take any step numbers~\citep{nichol2021improved, dhariwal2021diffusion} and we set the step numbers as 60, 80, 100, 120, and 140. For a fair comparison, baseline algorithms based on adjoint sensitivity analysis were evaluated using the same step numbers as our algorithm.

\noindent \textbf{Structural characteristics. }
The feature length is set to three distinct values: 0.224, 0.562, and 0.895 $\mu m$, corresponding to Gaussian filter standard deviations of $\sigma = 2$, $\sigma = 5$, and $\sigma = 8$, respectively. The derivation of these values is detailed in Supporting Information (S4). Additionally, we characterize the generated structures by quantifying the number of “islands,” which represent clusters of high-refractive-index regions, and evaluating their Minimum Feature Size (MFS). A lower number of islands is preferred for fabrication, as it reduces structural fragmentation, while a larger MFS is desirable to ensure manufacturability.

\subsection{Problem setup 1: bending waveguide}
\label{PS1}
Our first target problem involves the inverse design of a bending waveguide, as shown in Figure~\ref{problem setup} (a). The goal is to guide electromagnetic waves along a bent path with minimal loss. We employ silicon (Si) with a relative permittivity of $\varepsilon_r = 11.6$ and silicon dioxide (SiO$_2$) with $\varepsilon_r = 2.1$ as the constituent materials in a $10 \times 10 \mu\text{m}^2$ design region discretized into $64 \times 64$ pixels. The incident field is a Gaussian wave centered at a wavelength of $1.55 \mu\text{m}$.

To quantify performance, we define the figure of merit (FoM) as the conversion efficiency, given by

\begin{equation}
\text{FoM} = \text{Efficiency} = \frac{I_{\text{output}}}{I_{\text{input}}} 
= \frac{\left| \displaystyle \int_{\mathcal{C}} \mathbf{E} \times \mathbf{H}^* \cdot d\mathbf{l} \right|_{\text{output}}}{\left| \displaystyle \int_{\mathcal{C}} \mathbf{E} \times \mathbf{H}^* \cdot d\mathbf{l} \right|_{\text{input}}},
\end{equation}

where $I_{\text{output}}$ and $I_{\text{input}}$ denote the output and input intensities, respectively, computed from the line integral of the complex Poynting vector $\mathbf{E} \times \mathbf{H}^*$ along the path $\mathcal{C}$. Here, $\mathbf{E}$ and $\mathbf{H}$ are the electric and magnetic fields, and $\mathcal{C}$ denotes the integration contour.
 During the optimization, the material values in the design region are constrained to remain within the range of SiO$_2$ (represented as 0) and Si (represented as 1), ensuring that intermediate permittivity values do not exceed the physically meaningful limits.

 \begin{figure}[t]
  \centering
  \includegraphics[width=\textwidth]{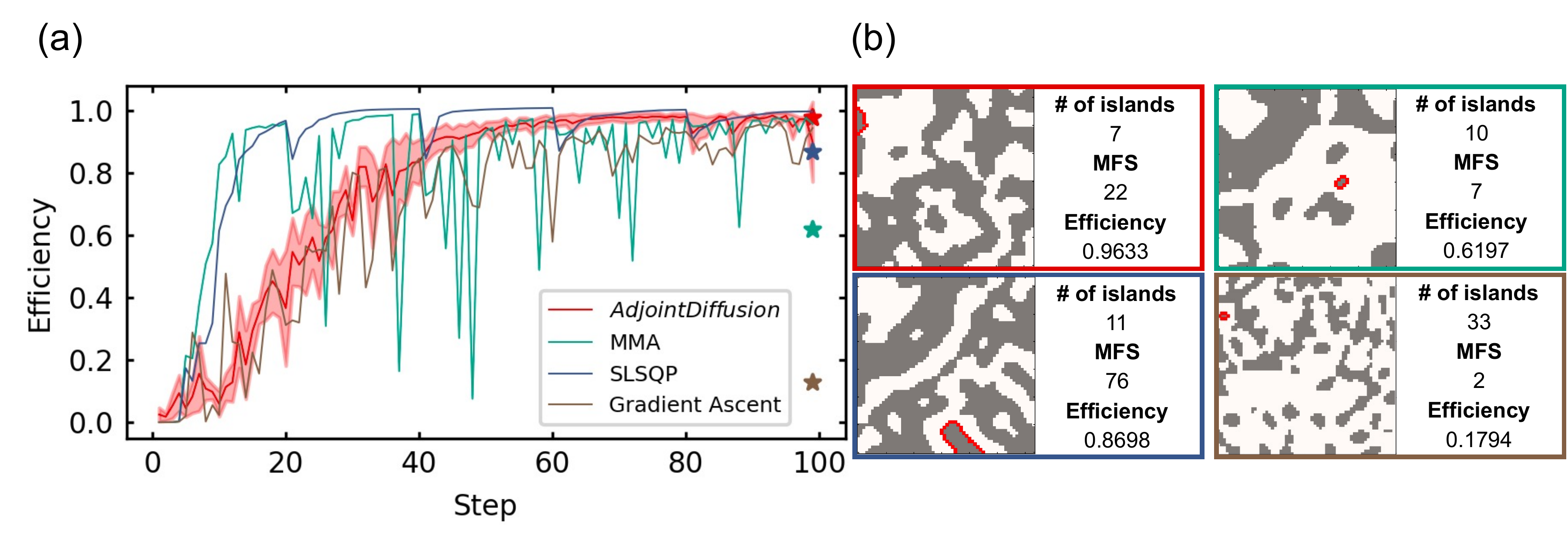}
  \caption{
  Optimization trajectories and generated structures of waveguide designs under the same fabrication conditions. (a) Bending efficiency over 100 steps of the optimization process for each algorithm. For \textit{AdjointDiffusion}, the red line denotes the mean efficiency across three experimental seeds, with the shaded region indicating standard deviation. The efficiency after post-processing is marked with a star symbol. (b) Generated structures with corresponding characteristics, where Minimum Feature Size (MFS) components are highlighted in red. 
  }
  \label{fig:wg}
  \end{figure}

  \begin{figure}[t]
    \centering
    \includegraphics[width=\textwidth]{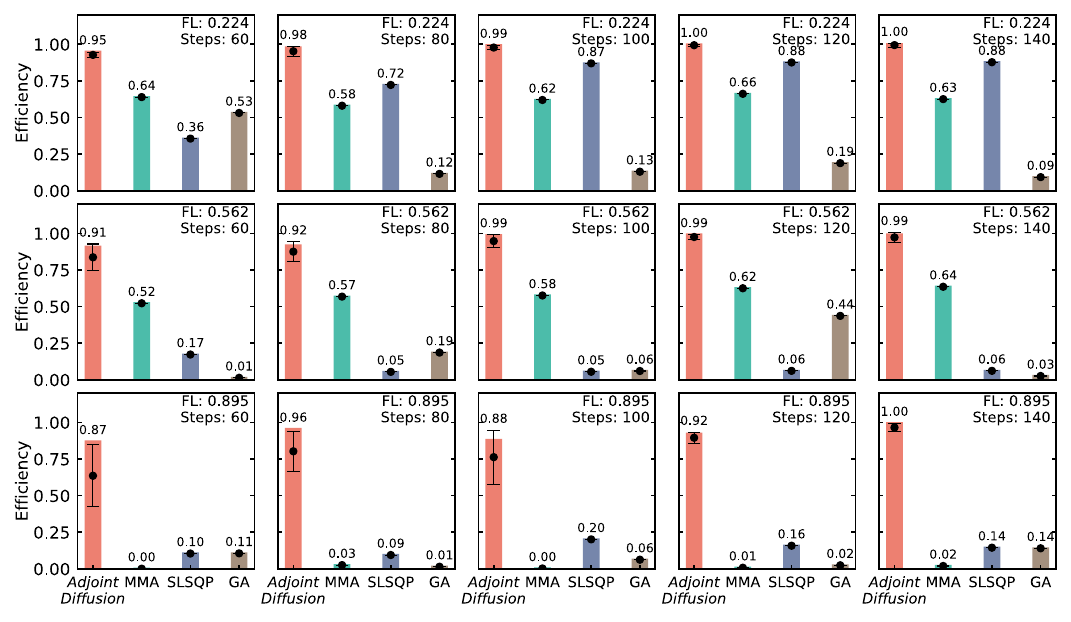}
    \caption{
    Maximum FoM (conversion efficiency) of generated waveguide structures across varying feature lengths (FL) and optimization steps. Three distinct feature length values are investigated: 0.224 (top row), 0.562 (middle row), and 0.895 (bottom row). Note that GA refers to the standard Gradient Ascent algorithm.
    }
    \label{fig_bar}
    \end{figure}

\subsection{Problem setup 2: Color routing-based CMOS image sensor}
\label{PS2}
The second example, illustrated in Figure~\ref{problem setup} (b), focuses on a color-routing-based CMOS image sensor operating in the visible spectrum.
Specifically, we aim to route red, green, and blue (RGB) light to different spatial locations to enable color separation directly at the pixel level. To achieve this, we introduce red (650nm), green (550nm), and blue (450nm) light as spatially overlapping, vertically incident beams from the top of the structure. Each color channel is represented by a Gaussian-shaped pulse centered at its respective wavelength, with a 40\% relative bandwidth to ensure sufficient spectral coverage. The incident beam uniformly spans the horizontal extent of the design region and is polarized along the out-of-plane direction, stimulating in-plane electric field responses. 
 For this design, we employ silicon nitride (Si$_3$N$_4$) with $\varepsilon_r = 4$ and silicon dioxide (SiO$_2$) with $\varepsilon_r = 2.1$ as the available materials. Like the problem setup 1, the design region consists of a $64 \times 64$ pixel grid. By tailoring the material distribution, the device directs each wavelength band (red, green, and blue) to a separate output waveguide, thereby achieving wavelength-selective routing.
 \begin{figure}[t]
  \centering
  \includegraphics[width=\textwidth]{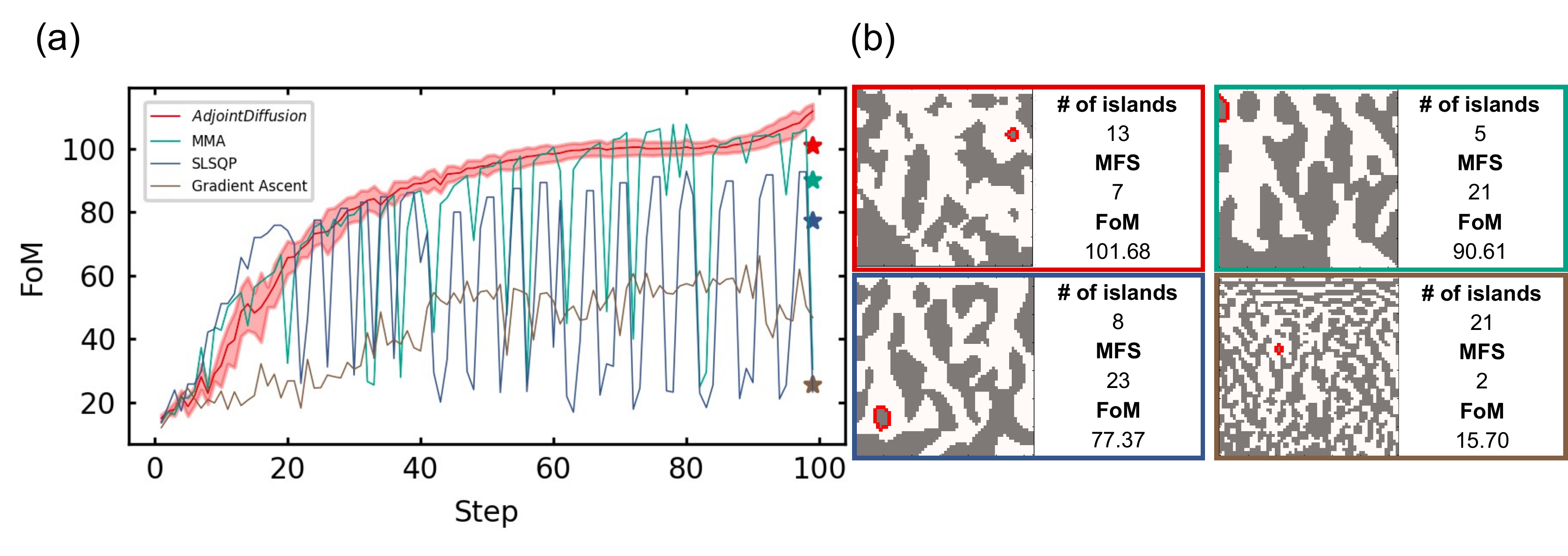}
  \caption{
  Optimization trajectories and generated structures of color router. The settings and baseline approaches are identical to those in Figure~\ref{fig:wg}, except that the optimization is performed for CMOS image sensors instead of waveguide structures.
  }
  \label{cis_optimization}
  \end{figure}

Initially, the FoM for each wavelength is computed via direct simulation, where a Gaussian source is incident from the top of the design region. The FoM is then formulated as:

\begin{align} \label{eq:1}
\displaystyle \text{FoM} = \sum_{\lambda \in \Lambda_R} \left\vert \textbf{\textit{E}}(\textbf{x}_{\scriptscriptstyle \textrm{R}},\lambda) \right\vert^2 +
\sum_{\lambda \in \Lambda_G} \left\vert \textbf{\textit{E}}(\textbf{x}_{\scriptscriptstyle \textrm{G}},\lambda) \right\vert^2 +
\sum_{\lambda \in \Lambda_B} \left\vert \textbf{\textit{E}}(\textbf{x}_{\scriptscriptstyle \textrm{B}},\lambda) \right\vert^2
\end{align}

where $\textbf{x}_{\scriptscriptstyle \textrm{R}}, \textbf{x}_{\scriptscriptstyle \textrm{G}}, \textbf{x}_{\scriptscriptstyle \textrm{B}}$ denote the central positions of the photodetectors in each subpixel, 
while $\textbf{\textit{E}}$ represents the electric field. The visible spectrum is divided into three wavelength bands: 400–500 nm for blue ($\Lambda_B$), 500–600 nm for green ($\Lambda_G$), and 600–700 nm for red ($\Lambda_R$). Within each spectral band, field intensities are sampled at discrete wavelengths, and the total efficiency is computed by summing the intensity contributions across these wavelengths.

\begin{figure}[t]
  \centering
  \includegraphics[width=\textwidth]{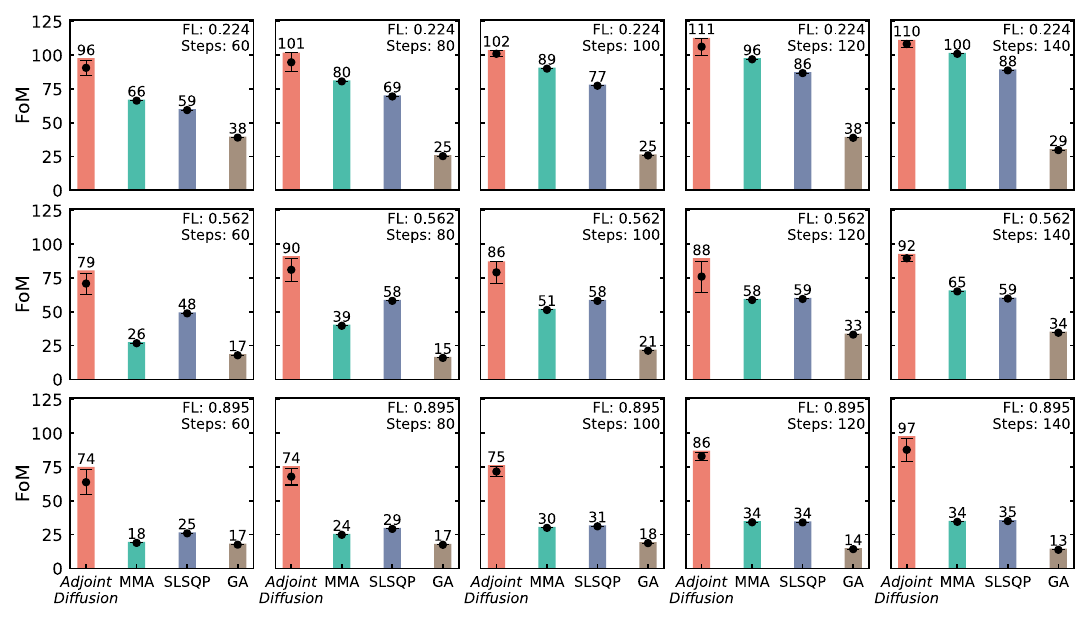}
  \caption{
  Maximum FoM of generated color router structures across varying feature lengths (FL) and optimization steps. As in Figure~\ref{fig_bar}, we consider three different feature length values: 0.224 (top row), 0.562 (middle row), and 0.895 (bottom row). Note that the definition of the FoM in the problem setup, which is based on the sum of light intensities, differs from the conversion efficiency used in Figure~\ref{fig_bar}. Additionally, GA refers to the standard Gradient Ascent algorithm.
  }
  \label{fig_bar_cis}
  \end{figure}

\subsection{Optimization performance across algorithms}

We present the optimization process and results of our algorithms compared to the baseline algorithms for two target problems. Figure~\ref{fig:wg} (a) and Figure~\ref{cis_optimization} (a) illustrate the optimization process of multiple algorithms optimizing bending waveguide and color router, respectively. To ensure a fair comparison, all methods were run for the same number of iterations (100), with a total of 200 simulations performed for each approach. While all methods utilize adjoint gradients, their application varies: MMA and SLSQP employ nonlinear optimization techniques, \textit{AdjointDiffusion} applies gradients within the reverse process of a diffusion model, and Gradient Ascent simply adds the gradient update. Note that baseline algorithms (MMA, SLSQP, and Gradient Ascent) require a meticulous binarization process and conic filters~\cite{kim2025freeform}. The star-shaped markers represent the efficiency after a direct threshold-based binarization at 0.5, with the red star (\textit{AdjointDiffusion}) indicating the averaged final result over three different random seeds.  The results demonstrate that \textit{AdjointDiffusion} not only discovers optimal structures but also exhibits robustness to the binarization process. This robustness stems from the fact that \textit{AdjointDiffusion} is trained on binary data, inherently guiding the generated structure towards binarity without requiring explicit binarization tuning. In contrast, baseline methods rely on scheduled binarization controlled by multiple hyperparameters, making them more sensitive to the choice of binarization strategy.
Figure~\ref{fig:wg} (b) and Figure~\ref{cis_optimization} (b) present the final structures obtained by each optimization algorithm, along with key structural characteristics such as the number of isolated high-refractive-index regions (islands) and the Minimum Feature Size (MFS). The red boundaries highlight the smallest MFS regions in each structure. While some algorithms may show slight advantages in structural characteristics, the overall differences are minor, except for Gradient Ascent, which produces highly fragmented and impractical designs. \textit{AdjointDiffusion} maintains a reasonable balance between the number of islands and MFS, producing a manufacturable structure without excessive fragmentation. As an example, we provide the optimized bending waveguide structure for a feature length of 0.224 and 100 optimization steps. For this design, \textit{AdjointDiffusion} achieved an average conversion efficiency of $98.00\%$ across three random seeds, significantly outperforming MMA (61.97\%), SLSQP (86.98\%), and Gradient Ascent (17.94\%).

In Figure~\ref{fig_bar} and Figure~\ref{fig_bar_cis}, we compare the optimization results of \textit{AdjointDiffusion} with multiple baseline algorithms across a range of design conditions (feature lengths of 0.224, 0.562, and 0.895 $\mu m$) and step numbers (60, 80, 100, 120, and 140). The results show that \textit{AdjointDiffusion} consistently outperforms nonlinear optimization methods and the standard Gradient Ascent (GA) algorithm under all tested scenarios. The error bars in the red graphs indicate standard deviations due to stochasticity, which remain relatively small and do not significantly impact the overall performance trends. Note that gradient-ascent-based baseline algorithms such as MMA, SLSQP, and standard Gradient Ascent are deterministic and therefore do not exhibit variance in their results.

\begin{figure}[t]
  \centering
  \includegraphics[width=\textwidth]{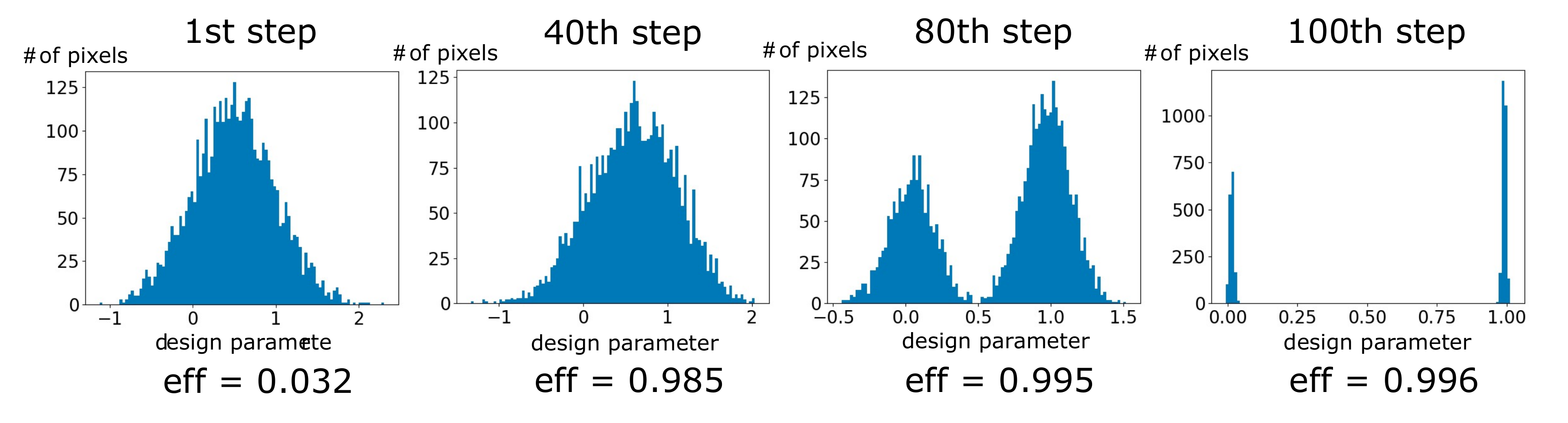}
  \caption{
  Pixel value (design parameter) distribution and efficiency (eff) of the intermediate structure $x_t$ at each step of the bending waveguide optimization process. Note that design parameters corresponding to low permittivity are represented as 0, and those corresponding to high permittivity are represented as 1. The initial structures are generated by sampling from a standard Gaussian distribution. Regions with negative design parameters, while non-physical, can emerge due to the unconstrained nature of the diffusion process.
  At the 40th step, the distribution is shifted to increase efficiency to 0.985. At the 80th step, the diffusion prior divides the distribution into two groups to binarize. Finally, at the 100th step, the structure is mostly binarized.
  }
  \label{pixeldist}
  \end{figure}

Figure~\ref{pixeldist} shows the histogram of pixel values for the generated structure $\bs{\varepsilon}_t$ at each step $t$. The results suggest that the binary prior of our diffusion model naturally promotes binarization, and this effect works effectively in conjunction with the proposed guidance approach in~\eqref{eq:guided_sampling}. As previously discussed, this natural binarization is a direct consequence of the manifold-constrained property introduced by our guidance term, encouraging generated structures to remain close to the binary data manifold.

\section{Conclusion}
We have presented \textit{AdjointDiffusion}, a physics‑guided inverse‑design framework that integrates adjoint sensitivity analysis with denoising diffusion models to rapidly generate high‑performance, fabrication‑compliant photonic structures. By training on a synthetic binary dataset and injecting adjoint gradients directly into the reverse‑diffusion process, our approach obviates complex binarization steps and reduces the required simulation budget to just $200$ runs, several orders of magnitude fewer than typical deep‑learning methods. We validated AdjointDiffusion on bent‑waveguide and CMOS color‑router design tasks, where it outperformed conventional adjoint‑based optimizers in both figure‑of‑merit and manufacturability. These results establish AdjointDiffusion as one of the most simulation‑efficient deep‑learning solutions for photonic inverse design, offering a streamlined pathway toward next‑generation photonic device development.

\section{Limitations and further works}
Though we have limited our target problem to photonic inverse design, our algorithm is not restricted to photonics. Since adjoint sensitivity analysis is a broadly applicable technique, our method can be extended to other domains such as mechanics~\cite{pappalardo2017adjoint} and fluid dynamics~\cite{mcnamara2004fluid}.
Also, while the application in this paper is limited to different feature sizes of the input structure, the applicability of our algorithm is unlimited; any structural constraints can be incorporated into the generation process by training on an image dataset of constrained structures. For example, our method can generate designs which adhere to sophisticated fabrication constraints such as in~\citet{schubert2022inverse}.
On top of that, the current application of our work is limited to two-dimensional structures. However, with advancements in 3D simulations and the development of 3D diffusion models, our approach can be extended to three-dimensional design optimization. This expansion would allow for the optimization of volumetric structures and will enable applications such as multi-layered lithography limitations or high-aspect-ratio manufacturability requirements.

\begin{acknowledgement}

The authors thank Jinwoong Cha for the valuable discussion on the fabrication constraints of devices. The authors also thank Jay S. Zou for the informative discussion on adjoint methods. 

\end{acknowledgement}

\begin{suppinfo}

The following files are available free of charge.
\begin{itemize}
  \item Supporting Information.pdf: introduction to background algorithms and additional results supporting our method
\end{itemize}

\subsection{source code}
We open our source code and simulation file freely accessible at \href{https://github.com/dongjin-seo2020/AdjointDiffusion}{https://github.com/dongjin-seo2020/AdjointDiffusion}

\end{suppinfo}


\providecommand{\latin}[1]{#1}
\makeatletter
\providecommand{\doi}
  {\begingroup\let\do\@makeother\dospecials
  \catcode`\{=1 \catcode`\}=2 \doi@aux}
\providecommand{\doi@aux}[1]{\endgroup\texttt{#1}}
\makeatother
\providecommand*\mcitethebibliography{\thebibliography}
\csname @ifundefined\endcsname{endmcitethebibliography}  {\let\endmcitethebibliography\endthebibliography}{}

\end{document}


\newpage
\tableofcontents

\newpage

\section{Background} 

Here, we introduce the two main components of our paper. Variables in bold denote matrices or vectors, while variables in regular font represent scalars. Note that our notation, where $\bveps$ represents the design parameter, $\mathbf{x}$ represents the design space, and $\mathbf{z}$ represents random noise, differs from the convention commonly used in articles on diffusion models.

\subsection{Adjoint optimization}

Adjoint optimization uses adjoint sensitivity analysis to compute the adjoint gradient and update each coordinate with the gradient value. If we take an optical lens as an example, the FoM for optimization is the intensity of the electric field ($\textbf{\textit{E}}^{\textrm{dir}}$) at a focal point (${\mathbf{x}_f}$) generated by an incident plane wave from below:

\begin{equation}
\text{FoM} = \frac{1}{2} |\textbf{\textit{E}}^{\textrm{dir}}(\mathbf{x}_f)|^2.
\end{equation} First, we define the structural parameter as a permittivity indicator ($\varepsilon$), where $\varepsilon = 1$ represents a material with higher permittivity (e.g., Si) and $\varepsilon = 0$ represents a material with lower permittivity (e.g., Air). For convenience, we refer to this value as permittivity. Initially, we set all coordinate permittivity values to 0.5 as the initial guess for the structure. Next, we determine the figure of merit (FoM) value through a direct simulation induced by an incident plane wave. In the direct simulation, the electric field at the target point $\textbf{\textit{E}}^{\textrm{dir}}(\mathbf{x}_f)$ is computed and subsequently used as amplitudes of the adjoint sources, while $\textbf{\textit{E}}^{\textrm{dir}}\left(\mathbf{x}^\prime\right)$ at the design area is saved for the purpose of adjoint gradient calculation.

The variation in $\text{FoM}$ by the transmitted electric field is 
\begin{equation}
    {\delta \text{FoM}} \approx \textrm{Re}\left[\overline{{\textbf{\textit{E}}^{\textrm{dir}}(\mathbf{x}_f)}}\delta \textbf{\textit{E}}^{\textrm{dir}}(\mathbf{x}_f)\right].
    \label{eqn:eq3}
\end{equation}

The variation of the electric field at point $\mathbf{x}_f$, caused by the adjustment in the permittivity of the design space, can be expressed as 
\begin{equation}
    \delta \textbf{\textit{E}}^{\textrm{dir}}(\mathbf{x}_f) = \textbf{G}(\mathbf{x}_f, \mathbf{x}') \textbf{P}^{\text{ind}}(\mathbf{x}') = \textbf{G}(\mathbf{x}_f, \mathbf{x}')\delta\bveps(\mathbf{x}')\textbf{\textit{E}}^{\textrm{dir}}(\mathbf{x}'),
    \label{eqn:eq4}
\end{equation}
where $\mathbf{x}$ and $\mathbf{x}'$ indicate the positions in the monitor and the design space, respectively. $\textbf{P}^{\text{ind}}(\mathbf{x}')$ indicates the polarization density,  which is induced by the variation of the permittivity $\delta\bveps(\mathbf{x}')$.  $\overleftrightarrow{\textbf{G}}(\mathbf{x}_f, \mathbf{x}')$ is a Green’s function which represents the electric field at the point $\mathbf{x}_f$ generated by the unit dipole at the point $\mathbf{x}'$. The formula for the change in FoM becomes

\begin{equation}
{\delta \text{FoM}} = \textrm{Re}\left[\textbf{G}(\mathbf{x}_f, \mathbf{x}')\overline{{\textbf{\textit{E}}^{\textrm{dir}}(\mathbf{x}_f)}}\delta\bveps(\mathbf{x}')\textbf{\textit{E}}^{\textrm{dir}}(\mathbf{x}')\right].
\end{equation} 
The adjoint field $ \textbf{\textit{E}}^{\textrm{adj}}$ can be expressed as
\begin{equation}
\textbf{\textit{E}}^{\textrm{adj}}(\mathbf{x}') = \textbf{G}(\mathbf{x}_f, \mathbf{x}')\overline{{\textbf{\textit{E}}^{\textrm{dir}}(\mathbf{x}_f)}}.
    \label{eqn:eq5}
\end{equation}

Subsequently, the amplitude of the adjoint dipole source thereby being $\overline{{\textbf{\textit{E}}^{\textrm{dir}}(\mathbf{x}_f)}}$. The adjoint dipole sources backpropagate through the designable region and generate $\textbf{\textit{E}}^{\textrm{adj}}\left(\mathbf{x}^\prime\right)$. Finally, the gradient of the FoM with respect to changes in $\bveps(\mathbf{x}^\prime)$ within the design region is computed as 
\begin{equation}
\frac{\delta {\text{FoM}}}{\delta\bveps(\mathbf{x}')}=\textrm{Re}[\textbf{\textit{E}}^{\textrm{dir}}(\mathbf{x}^\prime)\textbf{\textit{E}}^{\textrm{adj}}(\mathbf{x}^\prime)].
\end{equation}

Adjoint optimization uses the gradient value acquired from adjoint sensitivity analysis. This gradient indicates how the design parameters should be adjusted to improve FoM. Typically, this adjustment is performed using a gradient ascent algorithm with a learning rate $\eta$:

\begin{equation}
\bveps_{i+1}=\bveps_i+\eta \nabla_{\bveps_i} \operatorname{FoM}\left(\bveps_i\right).
\end{equation}

After each step, since the generated structures are in grayscale, a binarization process is required. Binarization is scheduled to operate less intensively at the beginning and more intensively towards the end. This process is governed by two factors, \(\beta\) and \({\bs \zeta}\):

\begin{equation}
\mathrm{\bveps}^{\prime}=\frac{\tanh \left(\beta * {\bs \zeta}\right)+\tanh \left(\beta *(\mathrm{\bveps}-{\bs \zeta})\right)}{\tanh \left(\beta * {\bs \zeta}\right)+\tanh \left(\beta *(\mathbf{1}-{\bs \zeta})\right)}
\end{equation}

\(\beta\) is a scalar that modulates the extent of binarization applied to the structures. Initially, \(\beta\) is set to \(\beta_{\text{init}} = 2\) and progressively increases by a factor of \(\beta_{\text{scale}} = 2\) over \texttt{num\_betas} steps:

\begin{equation}
\beta \leftarrow \beta \times \beta_{\text{scale}}
\end{equation}

This increasing \(\beta\) ensures that binarization becomes more pronounced as the optimization proceeds. The update interval is determined by \texttt{update\_factor}, meaning that \(\beta\) is updated every \texttt{update\_factor} optimization steps.

If \(\beta\) reaches a sufficiently large value, specifically when

\begin{equation}
\texttt{cur\_beta} \geq 2^{(\texttt{num\_betas} + 1)},
\end{equation}

the optimization effectively stops applying projection smoothing, treating the binarization as finalized. In this case, the objective function no longer depends on \(\beta\), and the number of evaluations is reduced to 1. Otherwise, the objective function continues to depend on the current \(\beta\), and the number of evaluations is set to \texttt{update\_factor}.

For the specific implementation details of adjoint optimization process, please refer to the code at \href{https://github.com/dongjin-seo2020/AdjointDiffusion/tree/main/baseline_algorithms}{https://github.com/dongjin-seo2020/AdjointDiffusion/tree/main/baseline\_algorithms}.


\noindent \textbf{Adjoint optimization with nonlinear optimization algorithm. }
Gradient ascent algorithms fall into local optima in nonconvex objective functions. To resolve this issue, researchers utilize nonlinear optimization algorithms that search the entire function space for the global optimum. From the open-source library named \textit{NLopt}, we import MMA (Method of Moving Asymptotes) and SLSQP (Sequential Least Squares Quadratic Programming) as baseline algorithms. Both are gradient-based local optimization algorithms that are generally used in adjoint optimization due to their robustness in handling complex constraints and non-convex problems.

1) MMA creates a local approximation using the gradient of the function along with a quadratic penalty term to ensure cautious approximations. The key idea is that the approximation is both convex and separable, making it easy to solve using a dual method. This solution provides a new candidate point that is then evaluated. If the approximations are conservative, the process restarts at the new point; if not, the penalty is increased, and the problem is re-optimized.

2) SLSQP is a sequential quadratic programming (SQP) algorithm designed to solve nonlinear optimization problems by breaking them down into simpler quadratic programming (QP) subproblems. It works by using a second-order Taylor expansion to approximate the objective function and linearize the constraints. The algorithm iteratively refines the solution until it meets the convergence criteria. If the criteria are not met, the process is repeated.

\subsection{Diffusion models} 

Diffusion models are latent variable models characterized by two fundamental stochastic processes: the forward reverse diffusion processes. The forward process is basically a Markov chain with a Gaussian transition, where data is gradually perturbed by Gaussian noise according to a variance schedule $\{{\beta_t}\}_{t=1}^T$: $q(\bveps_{t} | \bveps_{t-1}) \coloneqq {\cal N} ( \bveps_t ; \sqrt{1 - \beta_t} \bveps_{t-1}, \beta_t {\mathbf I})$ where $\{\bveps_t\}_{t=1}^T$ are latent variables with the same dimensionality as data $\bveps_0 \sim q(\bveps_0)$. One important property of the forward process is that it admits \emph{one-shot} sampling of $\bveps_t$ at any timestep $t \in \{1, \dots, T\}$: $q_{\alpha_t}(\bveps_t | \bveps_0) = {\cal N} (\bveps_t; \sqrt{\alpha_t} \bveps_0, (1- \alpha_t) {\mathbf I})$ where $\alpha_t \coloneqq \prod_{s=1}^t (1 - \beta_s)$. The variance schedule is designed to respect $\alpha_T \approx 0$ so that $\bveps_T \sim {\cal N}(\mathbf{0}, \mathbf{I})$.

The reverse process is another Markov Chain with learnable Gaussian transition $p_{\mathbf{\theta}}( \bveps_{t-1} | \bveps_t ) \coloneqq {\cal N}( \bveps_{t-1} ; \mathbf{\mu}_{\mathbf{\theta}}(\bveps_t, t), {\mathbf \Sigma }_{\mathbf \theta}( \bveps_t, t )  )$, where the variance is typically fixed, e.g., ${\mathbf \Sigma}_{\mathbf \theta} ( \bveps_t, t )  = \beta_t {\mathbf I}$. The mean is often parameterized by noise-prediction network ${\bs z}_{\bs \theta}(\bveps_t, t)$:
\begin{align}
    {\bs \mu}_{\bs \theta}(\bveps_t, t) = \frac{1}{\sqrt{1-\beta_t}}\left( \bveps_t - \frac{\beta_t}{\sqrt{1 - \alpha_t}} {\bs z}_{\bs \theta}(\bveps_t, t)  \right).
\end{align}
Here ${\bs z}_{\bs \theta}(\bveps_t, t)$ is constructed by a noise-matching training that learns to predict noise added on clean data $\bveps_0$:
\begin{align}
    \arg \min_{\bs \theta}  \mathbb{E}_{\bveps_0, {\bs z}, t} [ \| {\bs z} - {\bs z}_{\bs \theta} ( \sqrt{\alpha_t} \bveps_0 + \sqrt{1 - \alpha_t} {\bs z}, t  )  \|_2^2 ],
\end{align}
where ${\bs z} \sim {\cal N}({\bs 0}, {\bs I})$ and $t \sim \text{Unif}(1, T)$.

Given a pretrained noise prediction model, the generation can be done by starting from $\bveps_T \sim {\cal N}({\mathbf 0}, {\mathbf I})$ and iteratively going through the reverse process down to $\bveps_0$:
\begin{align}
\label{eq:ancestral}
    \bveps_{t-1} =  \frac{1}{\sqrt{1-\beta_t}}\left( \bveps_t - \frac{\beta_t}{\sqrt{1 - \alpha_t}} {\bs z}_{\bs \theta}(\bveps_t, t)  \right) + {\bs \Sigma }^{1/2}_{\bs \theta}( \bveps_t, t ) {\bs \xi}, \quad {\bs \xi} \sim {\cal N}({\mathbf 0}, {\mathbf I}).
\end{align}
This process, often called \emph{ancestral sampling}, is actually a discretized simulation of a stochastic differential equation that defines $\{p_{\bs \theta}(\bveps_t)\}_{t=0}^T$, which guarantees to sample from $p_{\bs \theta}(\bveps_0) \approx q(\bveps_0)$.

\noindent \textbf{Tweedie's denoising formula.} An additional benefit offered by diffusion models is that they provide an estimate of the posterior mean of a noisy instance $\bveps_t$. Specifically, the posterior mean $\mathbb{E}[\bveps_0 | \bveps_t]$ can be estimated using a pretrained diffusion models ${\bs z}_{\bs \theta}(\bveps_t, t)$:
\begin{align}
    \hat{\bveps}_0 (\bveps_t) \coloneqq \mathbb{E}[\bveps_0 | \bveps_t] = \frac{1}{\sqrt{\alpha_t}} \left( \bveps_t - \sqrt{1 - \alpha_t} {\bs z}_{\bs \theta}(\bveps_t, t) \right),
\end{align}
where $\hat{\bveps}_0(\bveps_t)$ represents an estimate of the posterior mean $\mathbb{E}[\bveps_0 | \bveps_t]$ provided by the diffusion model ${\bs z}_{\bs \theta}(\bveps_t, t)$. This is a direct application of Tweedie's formula~, since the noise prediction neural network ${\bs z}_{\bs \theta}(\bveps_t, t)$ is equivalent to the score function $\nabla_{\bveps_t} \log p_{\bs \theta}(\bveps_t)$, \ie, ${\bs z}_{\bs \theta}(\bveps_t, t) = -\sqrt{1 - \alpha_t} \nabla_{\bveps_t} \log p_{\bs \theta}(\bveps_t)$.

\newpage
\section{Implementation details of \textit{AdjointDiffusion}}
\label{AAC}

\begin{figure}[H]
\centering
\includegraphics[width=0.9\textwidth]{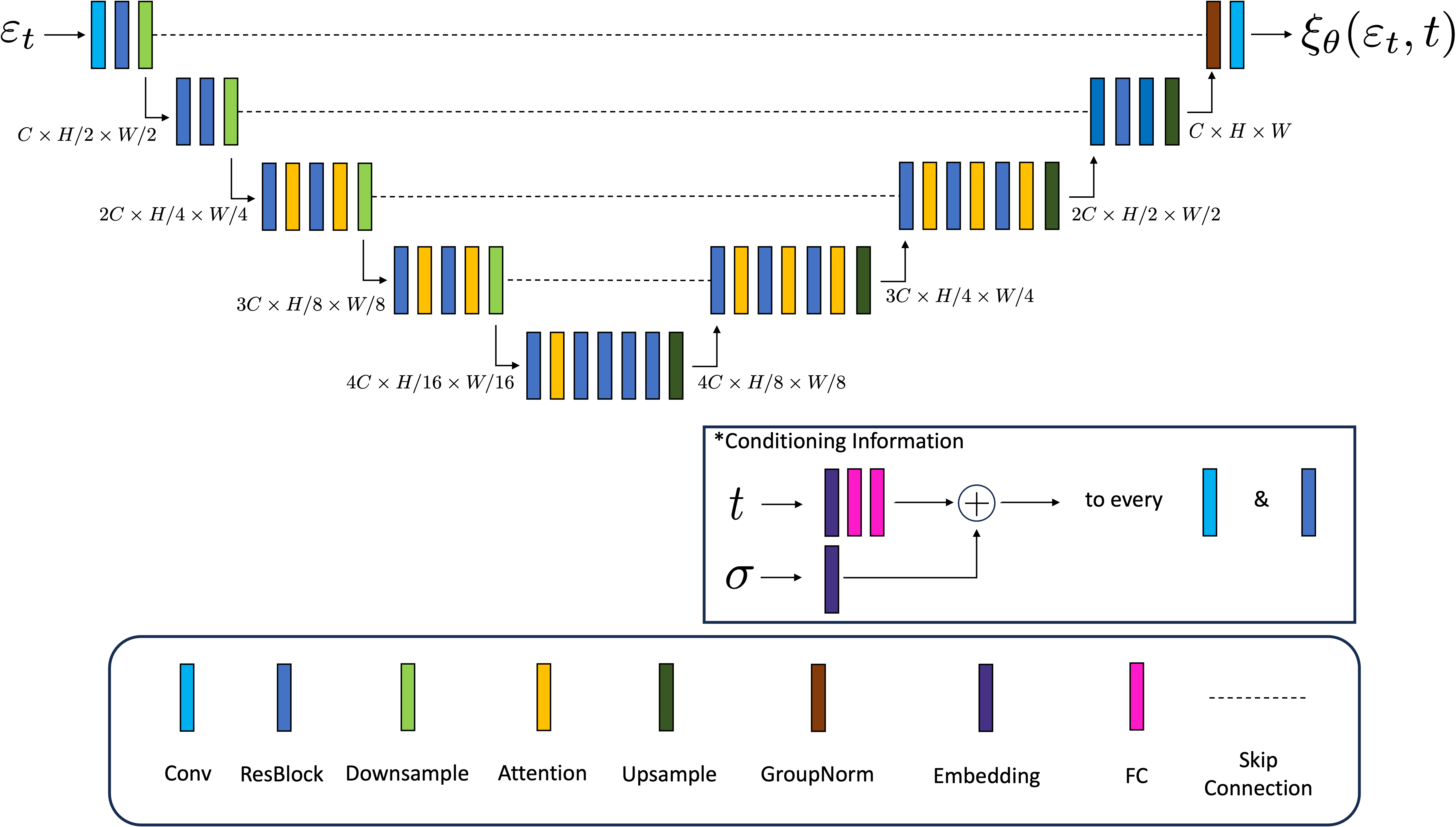}
  \caption{Architecture of the diffusion model employed in \textit{AdjointDiffusion}. Here, channel number $C = 128$ and spatial dimensions $H=
64$, $W = 64$. The model consists of multiple stages with convolutional layers (Conv), residual blocks (ResBlock), downsampling layers (Downsample), attention mechanisms (Attention), and upsampling layers (Upsample). The conditional information, consisting of time step $t$ and structural parameter $\sigma$, is embedded and added to each convolutional layer and residual block. The dimensions of the feature maps are progressively reduced and then increased, with skip connections linking corresponding layers across the downsampling and upsampling stages. Group normalization (GroupNorm) is applied to stabilize training.}
  \label{fig:arch}
\end{figure}

\noindent \textbf{Network architecture.} The neural network architecture we used for our diffusion backbone is based on U-Net. See Figure~\ref{fig:arch} for an overall illustration.

\noindent \textbf{Training setup.} The total number of timesteps is $T=1000$, and we use cosine noise schedule. For model training, we follow the recipe in and consider $L_{\text{hybrid}}$ to additionally learn the noise variance ${\bs \Sigma}_{\bs \theta}(\cdot, \cdot)$ for better performance when accelerating the sampling process. The training is performed with a learning rate of 1e-4, and the batch size is 128.

\noindent \textbf{Inference details.} We globally use $\eta = 1.0$ and 100 diffusion timesteps for sampling on both tasks (i.e., waveguide and CIS), while admitting variations whenever necessary (e.g., for ablation studies). We normalize the gradient (integrated in the reverse process) to have unit $l_{\infty}$ norm. More precisely, we employ $\nabla^*_{\bs{\varepsilon}_t} \operatorname{FoM}\left(\hat{\bs{\varepsilon}}_0\left(\bs{\varepsilon}_t\right)\right) \coloneqq \nabla_{\bs{\varepsilon}_t} \operatorname{FoM}\left(\hat{\bs{\varepsilon}}_0\left(\bs{\varepsilon}_t\right)\right) / \| \nabla_{\bs{\varepsilon}_t} \operatorname{FoM}\left(\hat{\bs{\varepsilon}}_0\left(\bs{\varepsilon}_t\right)\right) \|_{\infty}$ instead of $\nabla_{\bs{\varepsilon}_t} \operatorname{FoM}\left(\hat{\bs{\varepsilon}}_0\left(\bs{\varepsilon}_t\right)\right)$.

\noindent \textbf{Pseudocode.} We provide pseudocode which represents training and sampling processes of $\textit{AdjointDiffusion}$:
\algrenewcommand\algorithmicindent{0.5em}%
\begin{figure}[H]
    \begin{minipage}[t]{0.495\textwidth}
        \begin{algorithm}[H]
            \caption{Model training} \label{alg:training}
            \small
            \begin{algorithmic}[1]
            \vspace{.04in}
            \Repeat
                \State ${\bs \varepsilon}_0 \sim q({\bs \varepsilon}_0)$
                \State $t \sim \mathrm{Uniform}(\{1, \dotsc, T\})$
                \State $\sigma \sim \mathrm{Uniform}(\{2, 5, 8\})$ \Comment{{\scriptsize feature length para.}}
                \State ${\bs z} \sim \mathcal{N}({\bs 0}, {\bs I})$
                \State Take gradient descent step on
                \Statex $\qquad \nabla_{\bs \theta} \left\| {\bs z} - {\bs z}_{\bs \theta}(\sqrt{\alpha_t} {\bs \varepsilon}_0 + \sqrt{1-\alpha_t}{ \bs z}, t, \sigma) \right\|_2^2$
            \Until{converged}
            \vspace{.04in}
            \end{algorithmic}
        \end{algorithm}
    \end{minipage}
    \hfill
    \begin{minipage}[t]{0.495\textwidth}
        \begin{algorithm}[H]
            \caption{Physics-guided sampling} \label{alg:sampling}
            \small
            \begin{algorithmic}[1]
                \State{${\bs \varepsilon}_T \sim {\cal N}({\bs 0}, {\bs I})$}
                \For{$t \gets T$ to $1$}
                    \State{${\bs \xi} \sim {\cal N}({\bs 0}, {\bs I})$ if $t>1$, else ${\bs \xi} = {\bs 0}$}
                    \State{${\bs \varepsilon}_{t-1}^{\prime} \gets  \frac{1}{\sqrt{1 - \beta_t}}\big( {\bs \varepsilon}_t - \frac{\beta_t}{\sqrt{1-\alpha_t}} {\bs z}_{\bs \theta}({\bs \varepsilon}_t, t, \sigma) \big) + \sqrt{\beta_t} {\bs \xi}$}
                    \State{$\hat{\bs{\varepsilon}}_0 \gets \frac{1}{\sqrt{\alpha_t}} \left( \bveps_t - \sqrt{1 - \alpha_t} {\bs z}_{\bs \theta}({\bs \varepsilon}_t, t, \sigma) \right)   $}
                    \State{${\bs \varepsilon}_{t-1} \gets  {\bs \varepsilon}_{t-1}^{\prime} + \eta \nabla_{\bs{\varepsilon}_t} \operatorname{FoM}\left(\hat{\bs{\varepsilon}}_0\right)$}
                \EndFor
                \item[]
                \Return{${\bs \varepsilon}_0$}
            \end{algorithmic}
        \end{algorithm}
    \end{minipage}
\end{figure}

\noindent \textbf{Code implementation.} The implementation of our model is provided in \href{https://github.com/dongjin-seo2020/AdjointDiffusion}{https://github.com/dongjin-seo2020/AdjointDiffusion}.



\newpage
\section{Effect of step size $\eta$ on performance}
\label{AB}

\begin{figure}[H]
  \centering
  \includegraphics[width=\textwidth]{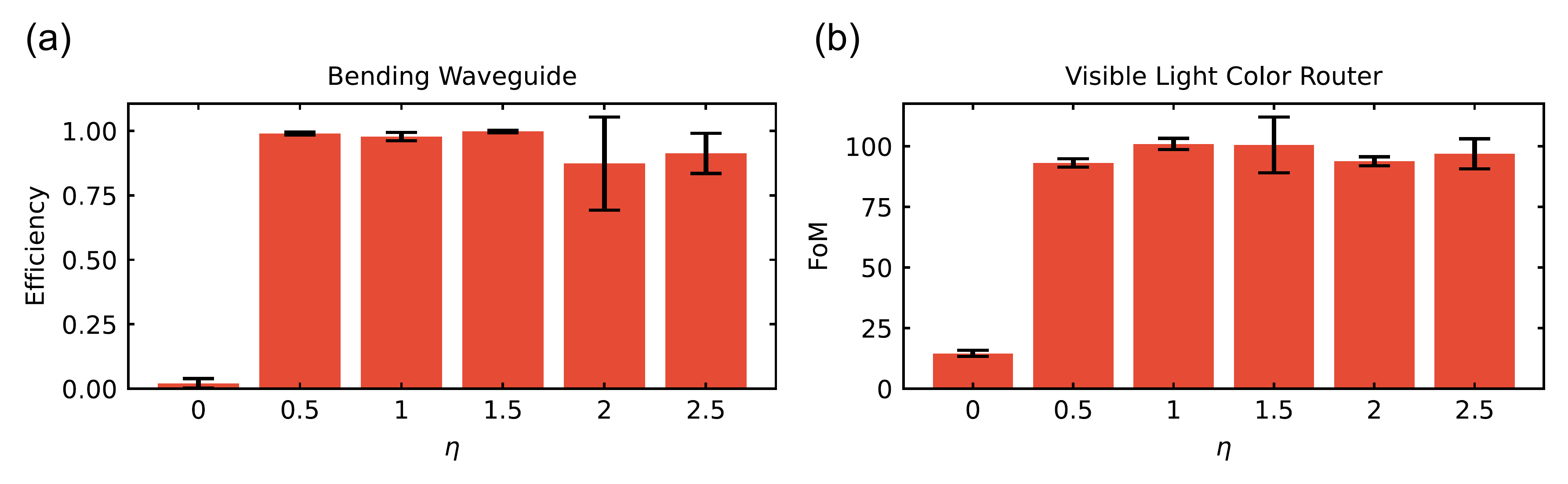}
  \caption{
  Efficiency of generated final structure from \textit{AdjointDiffusion}. Error bars represent the standard deviation of efficiency over three distinct experimental seeds.
  }
  \label{eta_on_performance}
  \end{figure}

Figure~\ref{eta_on_performance} illustrates the impact of varying the step size $\eta$, which is a coefficient multiplied to the additional adjoint gradient term in diffusion inference, on the performance of \textit{AdjointDiffusion}. At $\eta = 0$, the generation process is efficiency-agnostic, resulting in random generation based only on the datasets. For values such as $\eta = 0.5, 1, 1.5, 2 $ and $ 2.5$, optimization is applied to the structures. While performance may decrease when $\eta$ is 2 or higher, values around $\eta = 1$ show robustness of the algorithm performance, consistently approaching optimal efficiency (1.00 in the graph).

\newpage
\section{Calculation method of feature length in structure generation}
\label{FLCal}

\subsection{Feature length [pixels] from Gaussian filter (in image database)}
The Gaussian filter is a technique commonly used in image processing to smooth out and reduce noise. 
The minimum length of a structure that a Gaussian filter can resolve is related to the filter's standard deviation ($\sigma$), and the value determines the extent of blurring.

The full-width at half-maximum (FWHM) of the Gaussian function describes the effective spread of the filter. The FWHM is related to the standard deviation by the following formula:

\begin{equation}
\mathrm{FWHM}=2 \sqrt{2 \ln 2} \cdot \sigma
\label{FWHM}
\end{equation}
which calculates the width of the Gaussian function at half of its maximum value. The Feature Length, which we define as the minimum structure size that can be resolved by the Gaussian filter, is approximately equivalent to the FWHM.

\subsection{Feature length [$\mu$m] (in simulation)}
In Meep simulation, the Feature Length in micrometers [$\mu $m] is calculated as the Feature Length in pixels divided by the simulation resolution.
In our setup, we set the simulation resolution value as 21.

\subsection{Calculation}
We calculate the Feature Length of each dataset by applying Gaussian filter standard deviations $\sigma = 2, 5, 8$ to Equation~\ref{FWHM}, with the results presented in Table~\ref{FL}.

\begin{table}[h!]
\centering
\caption{Relationship between Gaussian filter standard deviation $\sigma$, Feature Length in pixels, and Feature Length in micrometers ($\mu$m).}
\begin{tabular}{ccc}
\hline
$\sigma$ & Feature Length [pixels] & Feature Length [$\mu$m] \\ \hline
2        &  4.7   & 0.224          \\ 
5        &  11.8  & 0.562          \\ 
8        &  18.8  & 0.895          \\
\end{tabular}
\label{FL}
\end{table}

\newpage
\subsection{Generated structures (bending waveguide)}

\begin{figure}[H]
  \centering
  \includegraphics[width=\textwidth]{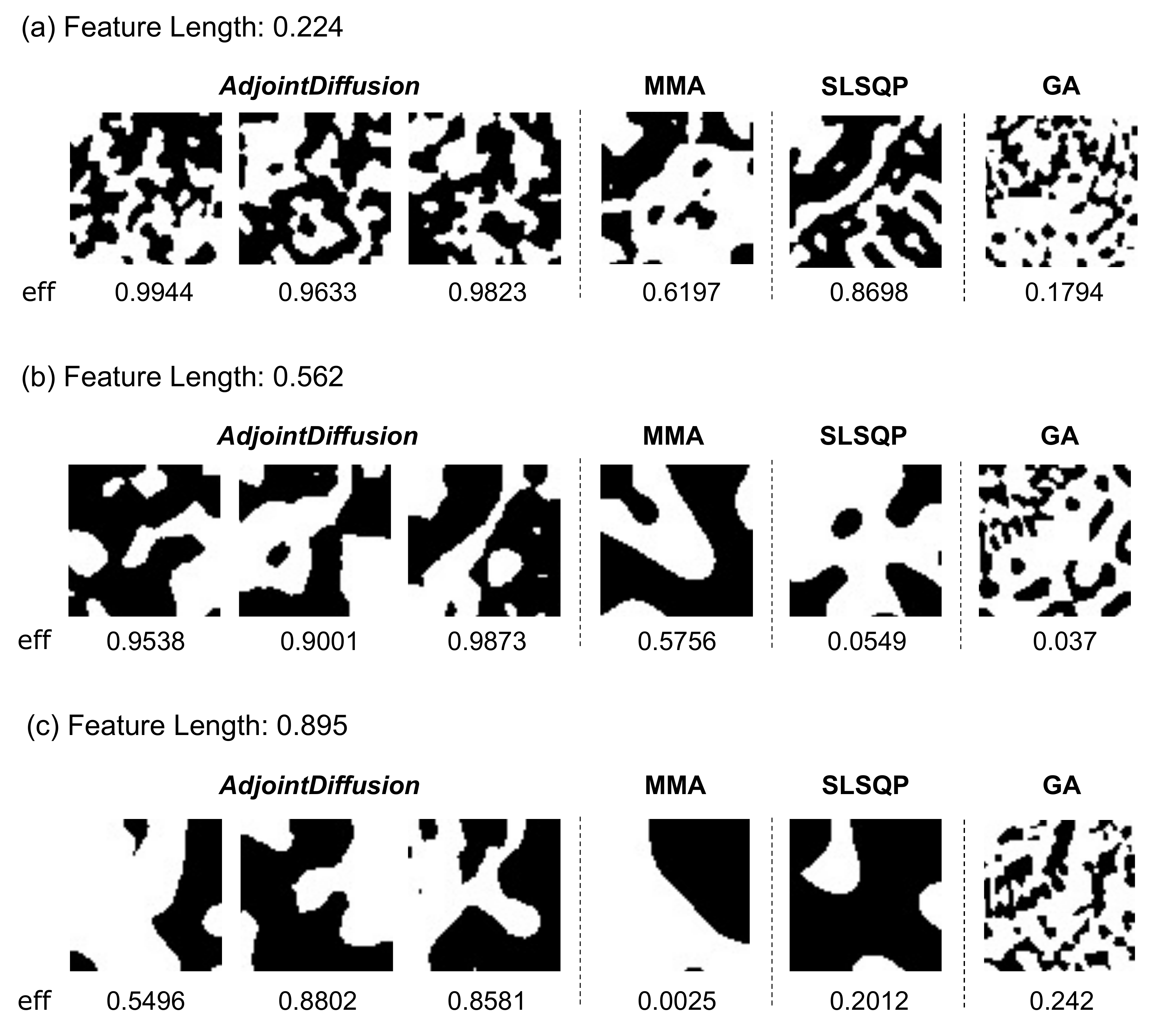}
  \caption{
  Generated structures for different Feature Lengths [$\mu$m] of (a) 0.224, (b) 0.562, and (c) 0.895 using various algorithms. The efficiency (eff) for each algorithm (\textit{AdjointDiffusion}, MMA, SLSQP, GA) is provided below the corresponding structures.
  }
  \label{GS}
  \end{figure}

We provide the generated structures from each algorithm. In Figure~\ref{GS}, the structures produced by \textit{AdjointDiffusion}, MMA, and SLSQP exhibit similar characteristic sizes. For the Gradient Ascent (GA) algorithm, we use the same simulation setup, but the Feature Length is not applied as expected, likely due to an issue with the Conic Filter. However, the smaller Feature Length does not disadvantage GA, so we report the results as they are.

\newpage
\section{FoM degradation from post-processing}
\label{BDAO}

We analyze the degradation of the Figure of Merit (FoM) following post-processing, which includes binarization and island deletion.

\subsection{Adjoint optimization}

Using MMA as an example, we observe that, while it achieves a high FoM during the optimization process, the algorithm is particularly vulnerable to post-processing steps, such as binarization and island deletion.

\begin{figure}[H]
  \centering
  \includegraphics[width=\textwidth]{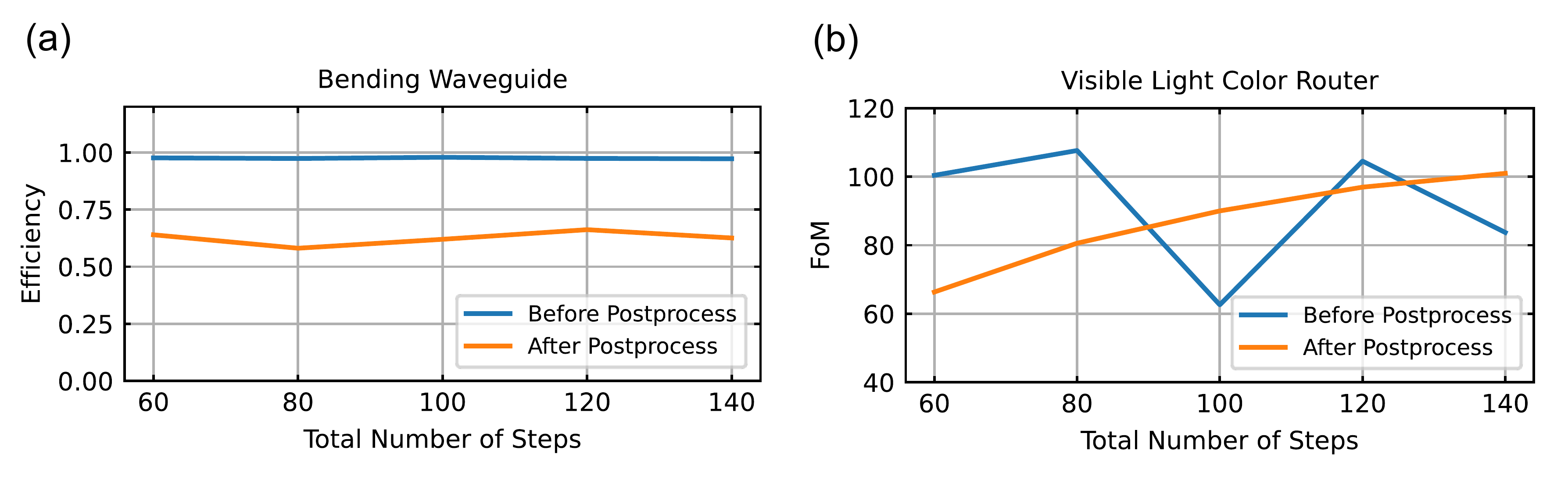}
  \caption{
  (a) Efficiency comparison before and after post-processing of MMA for the Bending Waveguide over various numbers of optimization steps. The blue line represents efficiency before post-processing, while the orange line represents efficiency after post-processing.
  }
  \label{fig:binarization_comparison}
  \end{figure}
  
  \begin{figure}[H]
  \centering
  \includegraphics[width=\textwidth]{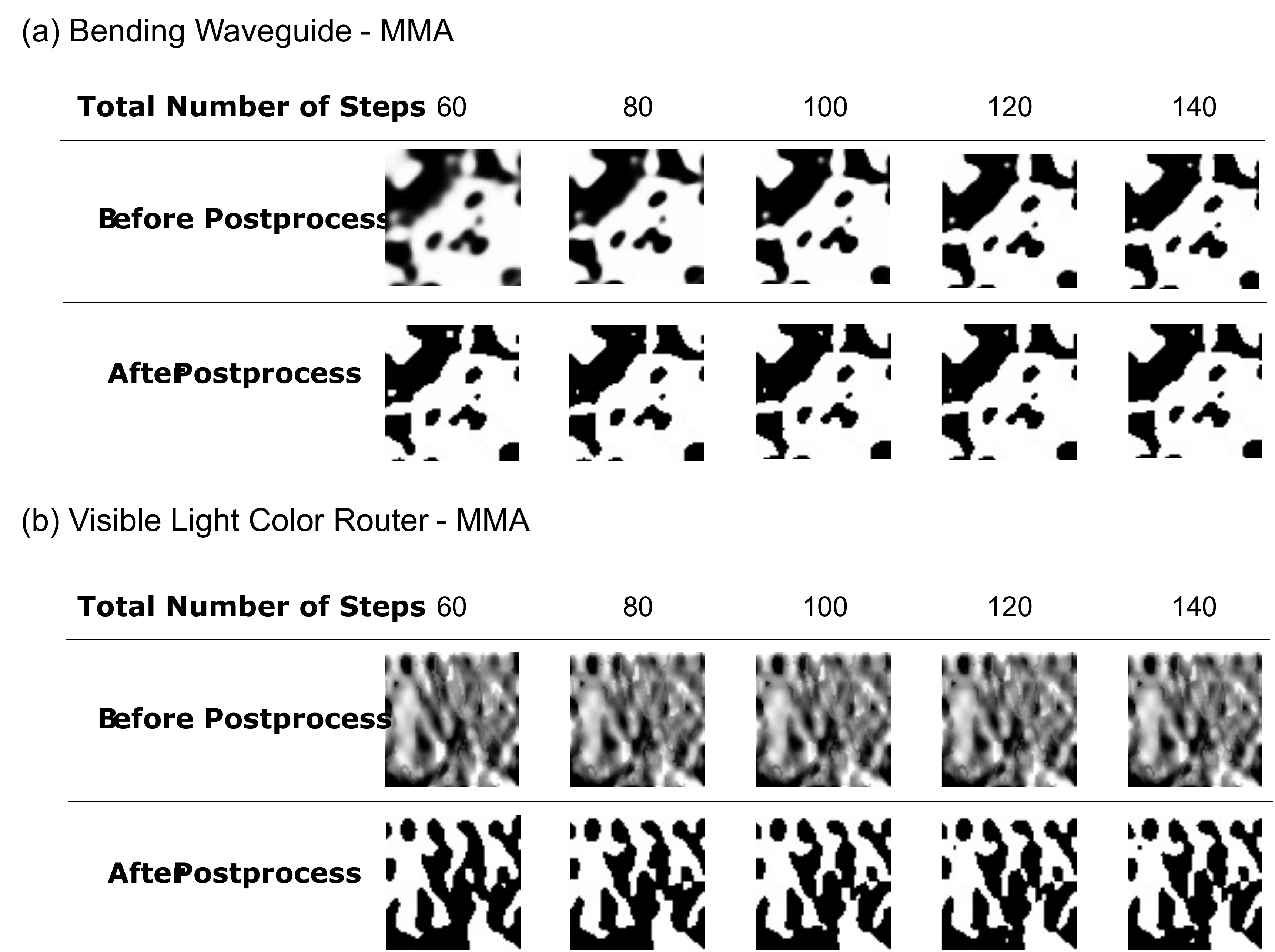}
  \caption{
  Visual comparison of generated structures before and after post-processing of MMA at different numbers of optimization steps. The top row shows the structures before post-processing, and the bottom row shows the corresponding structures after post-processing.
  }
  \label{fig:binarization_visual}
  \end{figure}


\newpage
\subsection{\textit{AdjointDiffusion}}

\textit{AdjointDiffusion} consistently remains robust to binarization and sometimes even demonstrates an unexpected improvement in FoM after post-processing.

\begin{figure}[H]
  \centering
  \includegraphics[width=\textwidth]{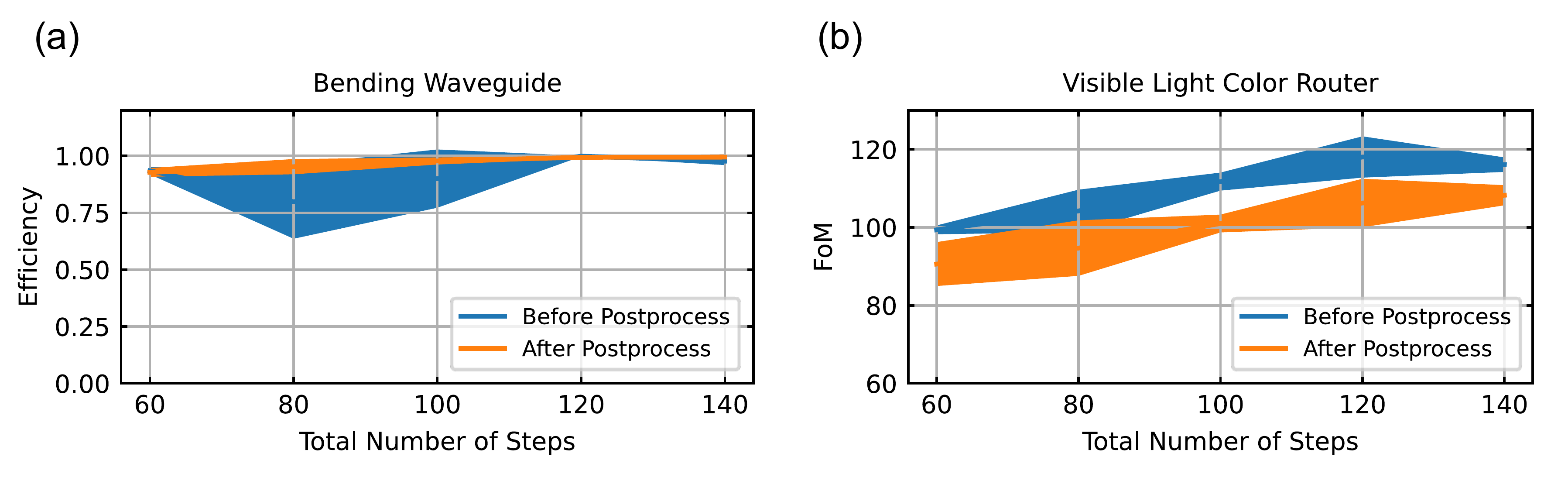}
  \caption{
  Efficiency comparison before and after post-processing of \textit{AdjointDiffusion} for the Bending Waveguide over various numbers of optimization steps. The blue region represents the mean and standard deviation of efficiency before post-processing, while the orange region represents the mean and standard deviation of efficiency after post-processing.
  }
  \label{fig:adjoint_binarization_comparison}
  \end{figure}
  
  \begin{figure}[H]
  \centering
  \includegraphics[width=\textwidth]{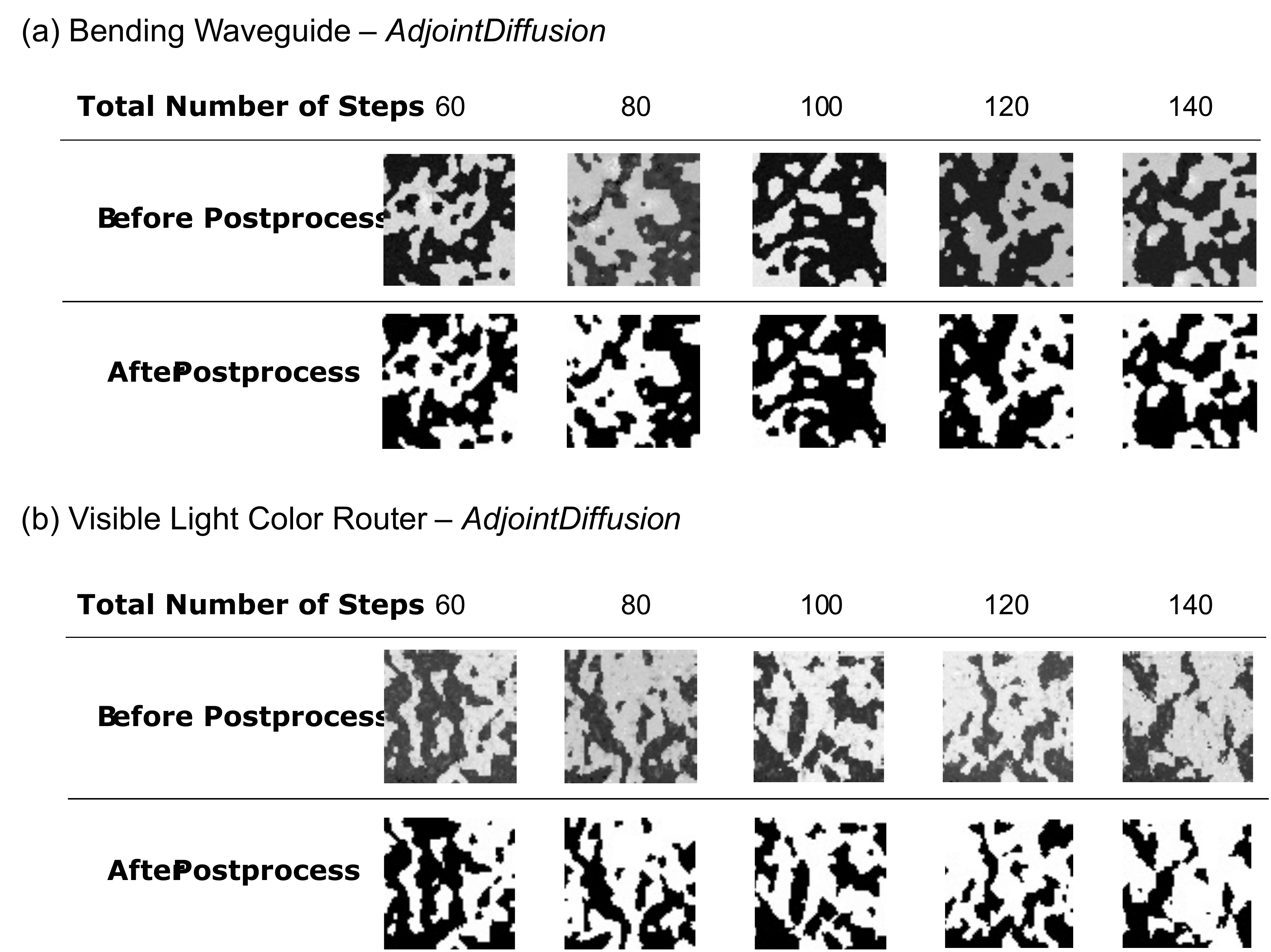}
  \caption{
  Visual comparison of generated structures before and after post-processing of \textit{AdjointDiffusion} at different numbers of optimization steps. The top row shows the structures before post-processing, and the bottom row shows the corresponding structures after post-processing.
  }
  \label{fig:adjoint_binarization_visual}
  \end{figure}

\newpage
\section{Visualization of simulation results}
\label{Sv}

In this section, we present the simulation results. The blue outlines represent the design boundaries, with the red line indicating the source and the blue lines indicating the electric field monitors.

\begin{figure}[H]
\centering
\includegraphics[width=\textwidth]{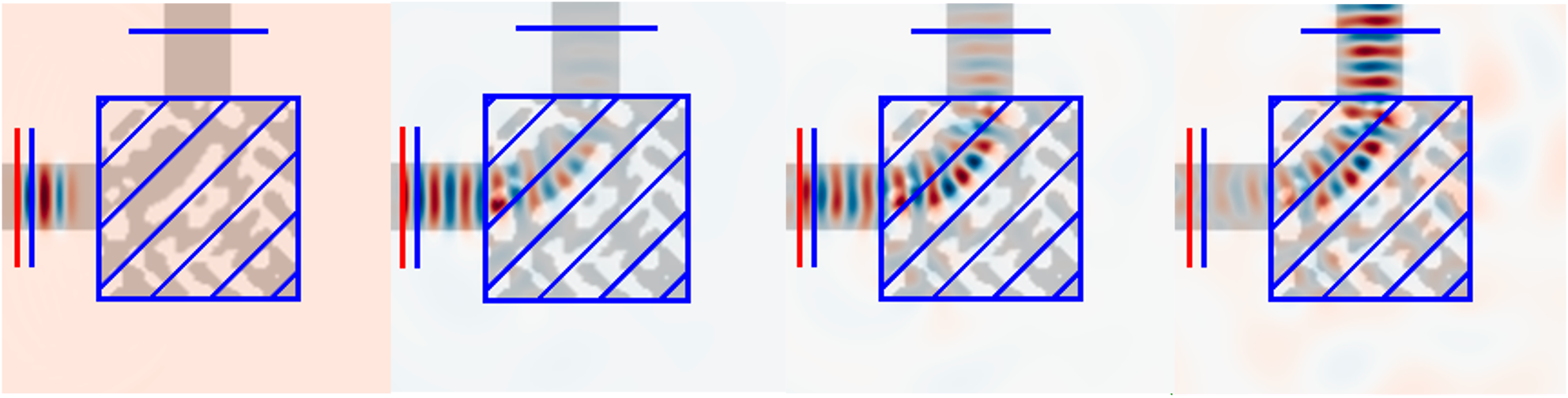}
  \caption{The evolution of the field distribution across different stages of the simulation}
  \label{sf1}
\end{figure}

Figure~\ref{sf1},~\ref{sf3}, and~\ref{sf5} provide a visualization of the simulation results, showing the evolution of the field distribution at various stages of the simulation. The color gradients illustrate the intensity and phase of wave propagation through the structure. The four parts of the figure correspond to specific points in the simulation timeline, capturing changes in the electromagnetic field as it interacts with the structure.

\subsection{Bending waveguide}

\subsubsection{Optimized structure from \textit{AdjointDiffusion}}

\subsubsection{Structures based on Human Intuition}

We additionally provide the visualized simulation results of two kinds of design based on human intuition.

\begin{figure}[H]
\centering
\includegraphics[width=0.25\textwidth]{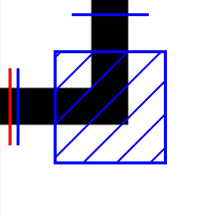}
  \caption{Right-angled waveguide}
  \label{sf2}
\end{figure}

\begin{figure}[H]
\centering
\includegraphics[width=\textwidth]{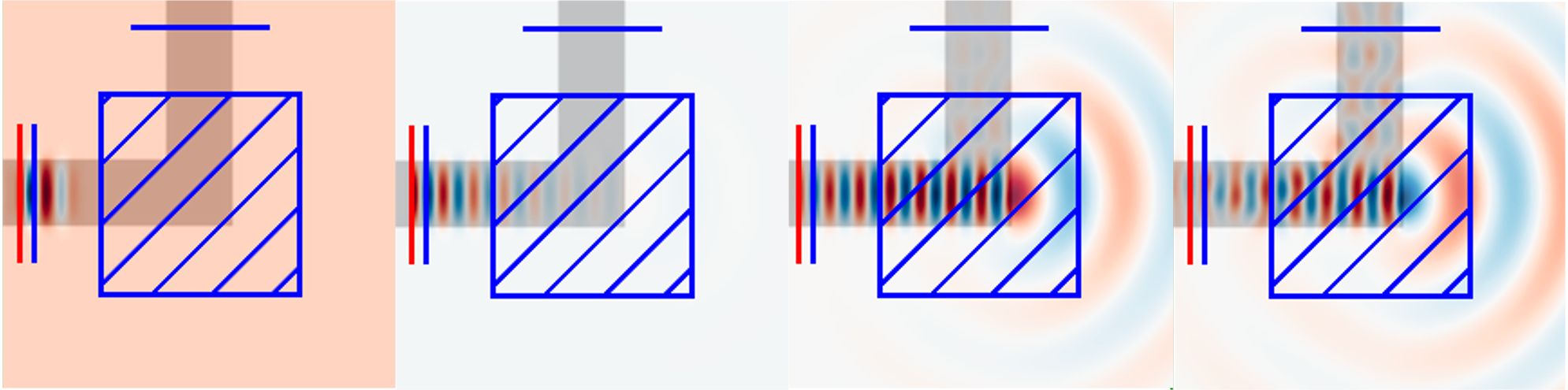}
  \caption{The evolution of the field distribution across different stages of the simulation for structure in Figure~\ref{sf2}. The efficiency of this structure is 0.0011.}
  \label{sf3}
\end{figure}

\begin{figure}[H]
\centering
\includegraphics[width=0.25\textwidth]{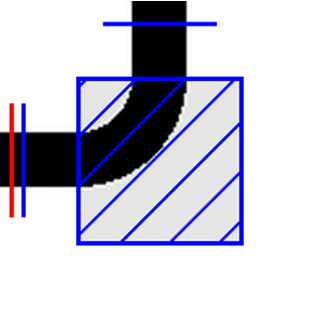}
  \caption{Curved waveguide}
  \label{sf4}
\end{figure}

\begin{figure}[H]
\centering
\includegraphics[width=\textwidth]{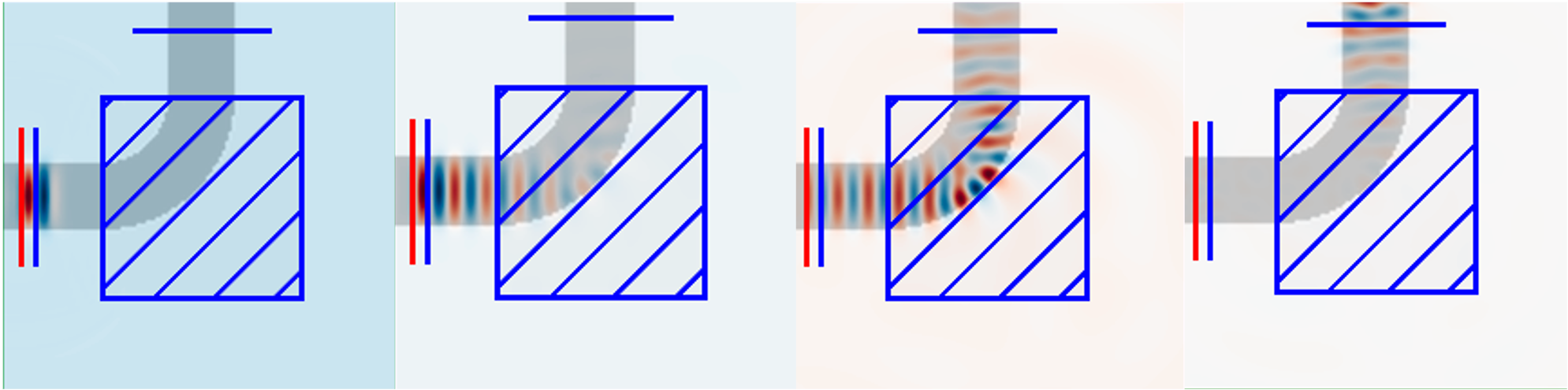}
  \caption{The evolution of the field distribution across different stages of the simulation for structure in Figure~\ref{sf4}. The efficiency of this structure is 0.8271.}
    \label{sf5}
\end{figure}

\subsection{Visible light color router}

\begin{figure}[H]
  \centering
  \includegraphics[width=\textwidth]{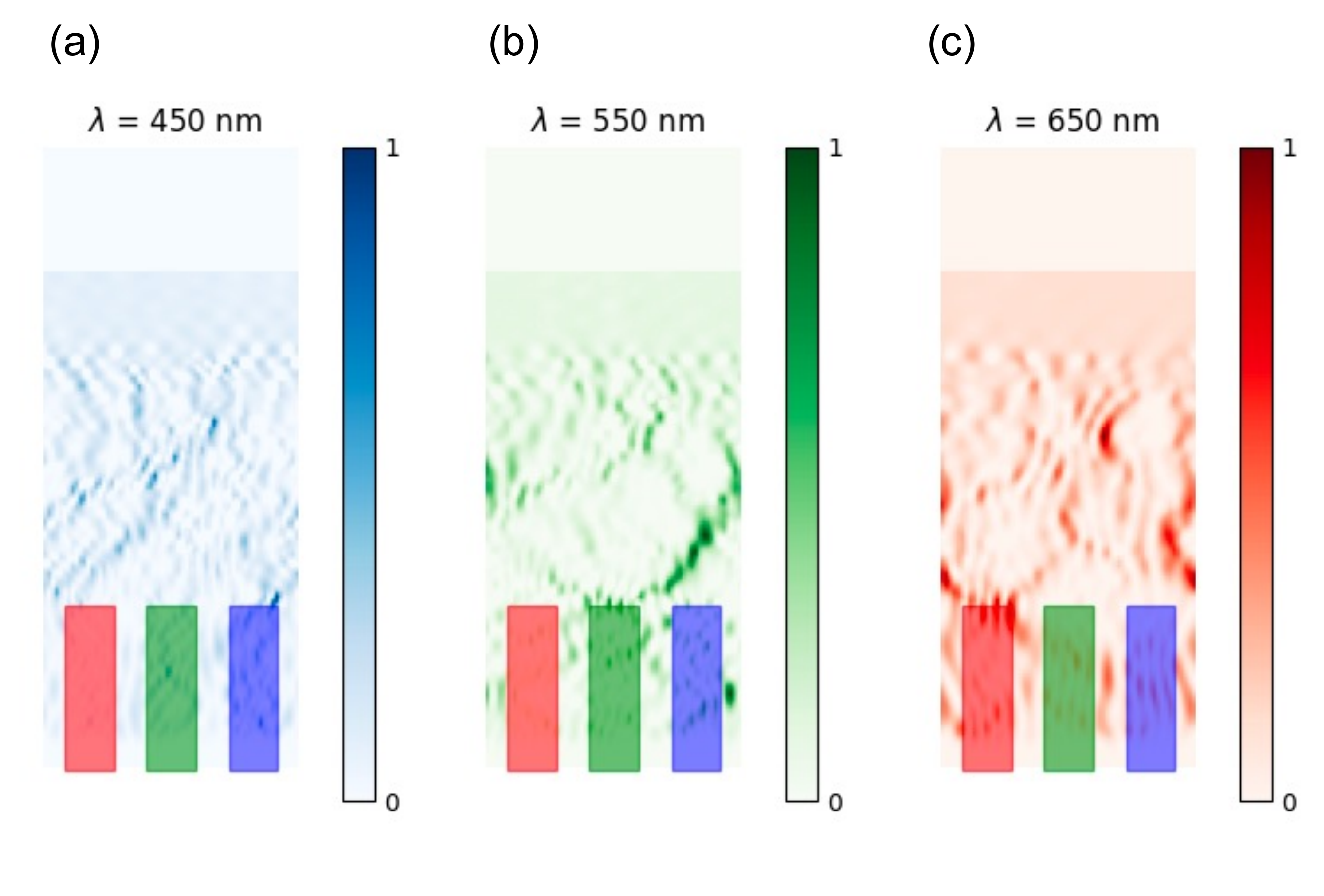}
  \caption{
  Intensity distributions of the optimized image sensor structure produced by our algorithm at three representative wavelengths: 450 nm, 550 nm, and 650 nm.
  }
  \label{fig:ColorRouterVis1}
  \end{figure}

\begin{figure}[H]
\centering
\includegraphics[width=0.7\textwidth]{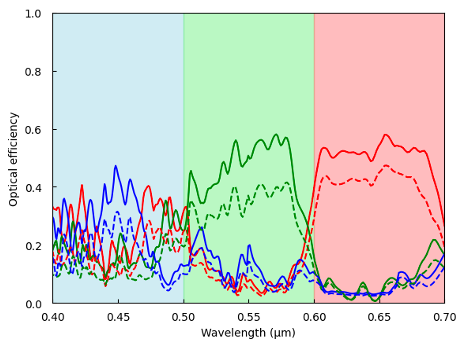}
  \caption{Optical efficiency of the three subpixels across the visible spectrum from the structure generated by our algorithm. Solid lines represent transmission-normalized efficiency, while dashed lines indicate absolute efficiency.}
    \label{dadad}
\end{figure}

\begin{figure}[H]
\centering
\includegraphics[width=\textwidth]{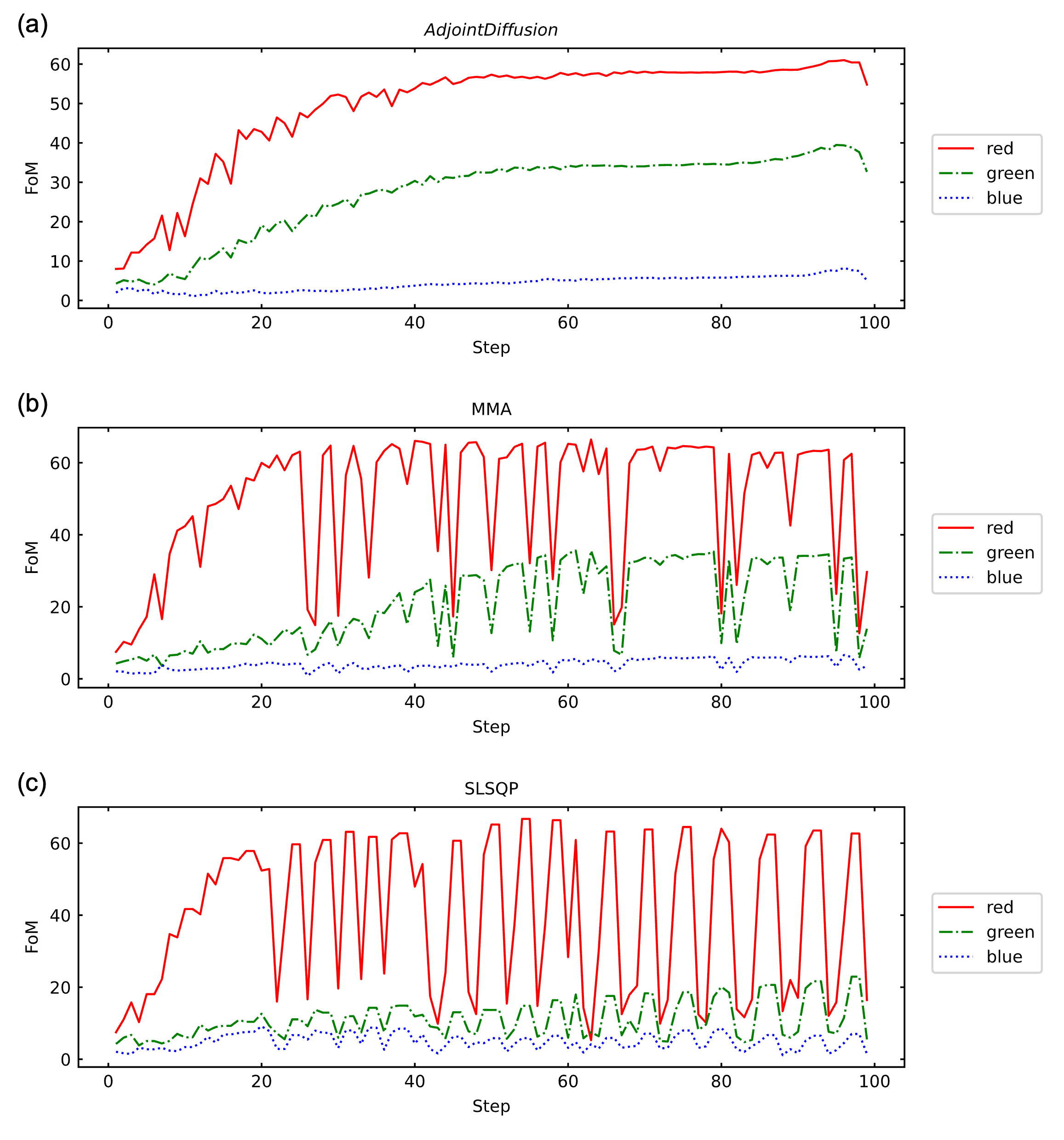}
    \caption{  
    Optimization trajectories of different algorithms for designing color router structures. (a) \emph{AdjointDiffusion} (ours), (b) MMA, and (c) SLSQP. The plots show the decomposed efficiency for three wavelength components: red, green, and blue. The red solid lines represent the efficiency contributions from red light, the green dash-dotted lines correspond to green light, and the blue dotted lines indicate the efficiency from blue light. The drops in FoM in each graph occur due to the binarization process. \emph{AdjointDiffusion} (a) experiences only a slight efficiency drop at the final stage, whereas MMA and SLSQP (b, c) exhibit large and frequent drops throughout the optimization.
    }
    \label{dadad2}
\end{figure}